\documentclass[lettersize,journal]{IEEEtran}
\usepackage{adjustbox}
\usepackage{amssymb}
\usepackage{amsmath}
\usepackage{cite}
\usepackage{url}
\usepackage{xcolor}
\usepackage{cite,graphicx,amsmath,amssymb}
\usepackage{subfigure}
\usepackage{citesort}
\usepackage{fancyhdr}
\usepackage{mdwmath}
\usepackage{mdwtab}
\usepackage{caption}
\usepackage{amsthm}
\usepackage{setspace}
\usepackage{algorithm}
\usepackage{algorithmic}
\usepackage{makecell}
\usepackage{diagbox}
\usepackage{balance} 
 \usepackage{multirow}
 \usepackage{booktabs}

\newtheorem{remark}{Remark}\newtheorem{theorem}{Theorem}

\newtheorem{lemma}{Lemma}

\newtheorem{corollary}{Corollary}

\allowdisplaybreaks
\makeatletter
\def\ScaleIfNeeded{
\ifdim\Gin@nat@width>\linewidth \linewidth \else \Gin@nat@width
\fi } \makeatother
 
\begin{document}
\title{Hybrid Active-Passive RIS Transmitter Enabled Energy-Efficient Multi-User Communications}
\author{Ao~Huang,~\IEEEmembership{Graduate Student Member,~IEEE,}
	Xidong~Mu,~\IEEEmembership{Member,~IEEE,}
	Li~Guo,~\IEEEmembership{Member,~IEEE,}
    and
	Guangyu~Zhu,~\IEEEmembership{Graduate Student Member,~IEEE}
	\thanks{Part of this work will be presented at the IEEE Wireless Communications and Networking Conference (WCNC), Dubai, United Arab Emirates, April 21-24, 2024~\cite{ao2024energy}.}
	\thanks{Ao Huang, Li Guo and Guangyu Zhu are with the Key Laboratory of Universal Wireless Communications, Ministry of Education, Beijing University of Posts and Telecommunications, Beijing 100876, China, also with the School of Artificial Intelligence, Beijing University of Posts and Telecommunications, Beijing 100876, China, and also with the National Engineering Research Center for Mobile Internet Security Technology, Beijing University of Posts and Telecommunications, Beijing 100876, China (email: huangao@bupt.edu.cn; guoli@bupt.edu.cn; zhugy@bupt.edu.cn).}
	\thanks{Xidong Mu is with the School of Electronic Engineering and Computer Science, Queen Mary University of London, London E1 4NS, U.K. (e-mail:xidong.mu@qmul.ac.uk).}}

\maketitle
\vspace{-1.5cm}
\begin{abstract}	
A novel hybrid active-passive reconfigurable intelligent surface (RIS) transmitter enabled downlink multi-user communication system is investigated. Specifically, RISs are exploited to serve as transmitter antennas, where each element can flexibly switch between active and passive modes to deliver information to multiple users. The system energy efficiency (EE) maximization problem is formulated by jointly optimizing the RIS element scheduling and beamforming coefficients, as well as the power allocation coefficients, subject to the user's individual rate requirement and the maximum RIS amplification power constraint. Using the Dinkelbach relaxation, the original mixed-integer nonlinear programming problem is transformed into a nonfractional optimization problem with a two-layer structure, which is solved by the alternating optimization approach. In particular, an exhaustive search method is proposed to determine the optimal operating mode for each RIS element. Then, the RIS beamforming and power allocation coefficients are properly designed in an alternating manner. To overcome the potentially high complexity caused by exhaustive searching, we further develop a joint RIS element mode and beamforming optimization scheme by exploiting the Big-M formulation technique. Numerical results validate that: 1) The proposed hybrid RIS scheme yields higher EE than the baseline multi-antenna schemes employing fully active/passive RIS or conventional radio frequency chains; 2) Both proposed algorithms are effective in improving the system performance, especially the latter can achieve precise design of RIS elements with low complexity; and 3) For a fixed-size hybrid RIS, maximum EE can be reaped by setting only a minority of elements to operate in the active mode.
\end{abstract}
\section{Introduction}
Reviewing the evolution of the fifth-generation (5G) networks and catering to the new challenges in future networks, the sixth-generation (6G) wireless applications will be further upgraded to achieve wider and deeper coverage with higher data rates and lower latency~\cite{al2015internet,saad2019vision}. In order to alleviate the current communication band resource constraints, the next generation of mobile communications is expected to evolve to higher frequency bands, e.g., millimeter wave (mmWave)~\cite{rangan2014millimeter} and terahertz (THz)~\cite{sarieddeen2021overview}. This will result in more severe attenuation of radio waves, posing greater challenges in terms of signal coverage as well as interference immunity for base stations (BSs). To circumvent this problem, high-band communications require that in addition to increasing the deployment density of BSs in the network, each BS has also to be equipped with more antennas~\cite{chaccour2022seven}. Massive multiple-input-multiple-output (MIMO) is a key technology in 5G~\cite{zhang2020prospective}, however, the large-scale deployment of BSs with massive antenna arrays may not be advisable for 6G networks. On the one hand, vast amounts of high-performance radio frequency (RF) chains are required for Massive MIMO, leading to a surge in hardware complexity, energy consumption, and cost. On the other hand, although the increased data rate of the system can be achieved by using spatial multiplexing techniques, the network performance could be severely damaged when the radio propagation environment exhibits poor scattering due to the lack of active modulation of the channels. Therefore, it is a priority to develop more advanced transceiver architectures with lower comprehensive costs for the future beyond 5G and 6G networks.

Recently, reconfigurable intelligent surfaces (RISs) have been envisaged as an advanced technology to realize system spectrum efficiency (SE) and energy efficiency (EE) improvement~\cite{wu2021intelligent}. Typically, an RIS comprises numerous tunable elements. By dynamically modulating the electromagnetic properties of these elements, RISs are able to customize favorable wireless propagation environments in a programmable manner, thereby facilitating the construction of ``Smart Radio Environments (SREs)''~\cite{di2020smart}. Compared to conventional Massive MIMO array antennas, RIS-based large-scale antennas do not require power-hungry RF chains, which significantly reduces the energy consumption and cost. In line with the above advantages, in addition to deploying RISs in networks to assist transmissions, RISs can also be designed as promising enablers of the transceivers for the upcoming 6G wireless communications~\cite{tang2020mimo,li2021beamforming}.
\subsection{Prior Works}
During the past years, a plethora of works in the literature were devoted to investigating RIS aided wireless systems, in which RISs usually act as full-duplex relays to assist signal transmission. For instance, the authors of~\cite{huang2019reconfigurable} studied the application of RISs in a downlink multiple-input single-output (MISO) system and maximized the system EE by the proper power allocation design. The authors of~\cite{mu2021capacity} characterized the fundamental capacity and rate regions of a RIS-assisted system, where different multiple access schemes were employed. The authors of~\cite{gan2021ris} performed the joint optimization of power control and phase shift in a RIS-assisted MISO network using statistical channel state information (CSI). Analytical formulations for the ergodic sum capacity are provided for both uplink and downlink scenarios. The authors of~\cite{xu2023reconfiguring} pointed out that RISs can serve not only as pure reflectors but also for transmitting additional information through phase modulation. The authors of~\cite{xu2023edge} discussed the potential application of RISs in edge learning and showed that deploying RISs in networks is particularly useful for distributed signal processing. The authors of~\cite{zhao2022cooperative} introduced a unified framework for joint timing and channel estimation in a communication network assisted by multiple RISs, where cooperative RIS reflection and synchronization design were taken into account. The authors of~\cite{huang2021multi} proposed a multi-hop RIS-aided transmission design to combat the propagation loss at TeraHertz-band frequencies. Specifically, deep reinforcement learning (DRL) is leveraged for hybrid beamforming design at the BS and multiple RISs. However, due to the near-passive nature, conventional RISs can only modify the phase of the incident signal, which significantly limits the achievable performance gains. To overcome the ``multiplicative fading'' effect of the signal, the concept of active RISs has emerged~\cite{zhang2022active}. With the support of extra power supplies, an active RIS is able to amplify the signal amplitude simultaneously while adapting the phase. Diverging from conventional practices of activating all elements, the authors of~\cite{xie2023to} presented a novel on-off mechanism for realizing flexible activation and deactivation of RIS elements.  Based on this, system EE in both passive and active RIS configurations was optimized. Moreover, a few studies have been conducted to unveil the benefits of utilizing hybrid active-passive RIS in network systems. Each element on the hybrid RIS operates in either an active mode or a passive mode. The authors of~\cite{nguyen2022hybrid} considered deploying a hybrid RIS in a MISO system to guarantee user fairness, the minimum rate among all users was maximized by jointly optimizing the transmit and RIS beamforming vectors. The authors of~\cite{kang2023active} investigated a worst-case user's ergodic capacity maximization problem for an active-passive RIS aided wireless network. The proposed RIS architecture comprises two co-located sub-surfaces, with one having all active reflecting elements and the other having all passive ones.

As a step forward, there is a growing research effort to study the design of RIS-based transceivers. Generally, with the ability to isolate the feed antenna and users on either side, thereby avoiding the collision of different electromagnetic (EM) waves, transmissive RISs are more suitable for use as transmitter enablers~\cite{li2021beamforming}. Moreover, compared to reflective RISs, transmissive RISs can support higher aperture efficiency and operating bandwidth~\cite{bai2020high}. Here, dynamic metasurface antenna (DMA) is a concept similar to RIS and has been proposed for aperture antenna design using configurable metamaterials~\cite{shlezinger2021dynamic}. A DMA-based array consists of parallel microstrips, each consisting of subwavelength and frequency-selective resonant metamaterial elements. Notably, DMA requires the same number of RF chains as microstrips, distinguishing it from RISs, which lack RF chains. In particular, the authors of~\cite{li2022joint} reported a new transmissive RIS enabled multi-tier computing framework, in which the joint beamforming design was explored with the aim of minimizing the overall system energy consumption. In another work~\cite{li2022robust}, the authors further studied the architecture of a downlink communication transceiver for simultaneous wireless information and power transfer (SWIPT) networks. Based on the imperfect CSI assumption, a robust system sum-rate maximization strategy was proposed. Very recently, the authors of~\cite{liu2023transmissive} proposed a transmissive RIS transmitter empowered cognitive radio network scheme to realize low-power, high-rate multi-stream transmissions.
\subsection{Motivations and Contributions}
As summarized above, prior research contributions have revealed that RIS is an extremely appealing technique for the future networks. Despite the burgeoning research, the pursuit of RIS-based transmitter design is still in its infancy. Since the incident signal can only be transmitted via conventional passive RIS without signal amplification, the achievable beamforming gain is greatly limited, even leading to the fact that the coverage capability of the BS may not be compatible with practical user communication requirements. This has led to our immense interest in multi-antenna transmitter design utilizing a hybrid active-passive RIS. Each element on the RIS is supposed to be able to adaptively switch between active and passive modes to cater for variable propagation environments. In this way, by imposing extra control circuits to RIS elements, not only can the signal processing capability of the transmitter be enhanced, but also higher beamforming gains can be achieved by fully reaping the degree-of-freedom (DoF) of signal amplitude modulation. To the best of our knowledge, the application of hybrid active-passive RIS has only been studied in a few relaying RIS assisted communications~\cite{nguyen2022hybrid,kang2023active}. However, the RIS optimization solutions proposed in both efforts are not applicable in this paper, where the RIS is used to construst a transmitter and the operating mode (i.e., active or passive) of each element has to be optimized. This provides the main motivation of this work to pursue a more precise design for each individual RIS element.

Bearing all these observations in mind, this paper makes a pioneering work to propose a novel transmitter architecture based on hybrid active-passive RIS. Recall that the thermal noise induced by active elements cannot be ignored~\cite{zhang2022active}, it is of paramount importance to strike a compromise between system performance and energy consumption. To address this issue, we aim to provide feasible and effective optimization algorithms to achieve energy-efficient communications in hybrid RIS transmitter enabled multi-user networks. The primary contributions are summarized as follows:
\begin{itemize}
\item 
We propose a hybrid RIS transmitter enabled downlink multi-user communication framework, where a hybrid active-passive RIS transmitter is exploited for information transmission. Based on the proposed transmitter architecture, we formulate an EE maximization problem by jointly optimizing the RIS element mode scheduling and beamforming vectors, as well as the power allocation coefficients, subject to the target rate requirement of individual users and the maximum amplification power limit of active RIS elements.  
\item 
We first invoke the Dinkelbach method to transform the original mixed-integer nonlinear programming (MINLP) problem into a more tractable form with a two-layer structure. Given the intractability of the problem, efficient algorithms are proposed based on alternating optimization (AO). Specifically, after deriving the maximum number of active elements a fixed-size RIS can accommodate, we apply an exhaustive search-based approach to determine the operating mode for each element. Then, the RIS beamforming and power allocation coefficients are alternately designed under a given RIS mode scheduling scheme. To further compress the computational complexity, we consider jointly optimizing the RIS element scheduling and beamforming coefficients by resorting to the Big-M formulation technique.
\item 
Our numerical results reveal that the proposed hybrid RIS transmitter achieves better performance compared to the baseline schemes relying on fully active/passive RIS or conventional RF chains. More DoFs introduced by mode switching can be properly exploited for higher beamforming gains at a slight cost of power consumption, thereby facilitating the improvements in EE performance. On top of that, for a fixed-size hybrid RIS, system EE is maximized with only a few elements operating in the active mode, which is a preferred setting in practical implementation.
\end{itemize}
\subsection{Organization and Notation}
The rest of this paper is organized as follows. Section II presents the transmission framework of the proposed hybrid active-passive RIS transmitter enabled downlink multi-user communication system and formulates an EE maximization problem. In Section III, an AO-based two-layer algorithm is proposed to solve the resulting non-convex optimization problem. Section IV develops a low-complexity joint hybrid RIS optimization scheme. Simulation results are provided in Section V, which is followed by conclusions in Section VI.\\
\indent \emph{Notations:} Scalars, vectors, and matrices are denoted by lower-case, bold-face lower-case, and bold-face upper-case letters, respectively. ${\mathbb{B}^{N \times M}}$ and ${\mathbb{C}^{N \times M}}$ denote the space of $N \times M$ matrices with binary and complex entries, respectively. ${{\mathbf{a}}^H}$ and $\left\| {\mathbf{a}} \right\|$ denote the conjugate transpose and the Euclidean norm of vector ${\mathbf{a}}$, respectively. ${\textup {diag}}\left( \mathbf{a} \right)$ denotes a diagonal matrix with the elements of vector ${\mathbf{a}}$ on the main diagonal. The distribution of a circularly symmetric complex Gaussian (CSCG) random variable with mean $\mu $ and variance ${\sigma ^2}$ is denoted by ${\mathcal{CN}}\left( {\mu,\sigma _k^2} \right)$. ${{\mathbf{1}}_{m \times n}}$ and ${{\mathbf{0}}_{m \times n}}$ denote the all-one and an all-zero matrices of size ${m \times n}$, respectively. ${\textup {Rank}}\left( \mathbf{X} \right)$ and ${\textup {Tr}}\left( \mathbf{X} \right)$ denote the rank and the trace of matrix $\mathbf{X}$, respectively. ${\textup {diag}}\left( \mathbf{X} \right)$ denotes a vector whose elements are extracted from the main diagonal elements of matrix $\mathbf{X}$. ${{\mathbf{X}}} \succeq 0$ indicates that $\mathbf{X}$ is a positive semidefinite matrix. ${\left\| {\mathbf{X}} \right\|_*}$ and ${\left\| {\mathbf{X}} \right\|_2}$ denote the nuclear norm and spectral norm of matrix $\mathbf{X}$, respectively.
\section{System Model and Problem Formulation}
\subsection{System Model}
\begin{figure}[t]
	\centering
	\setlength{\belowcaptionskip}{+0.1cm} 
	\includegraphics[width=	3.2in]{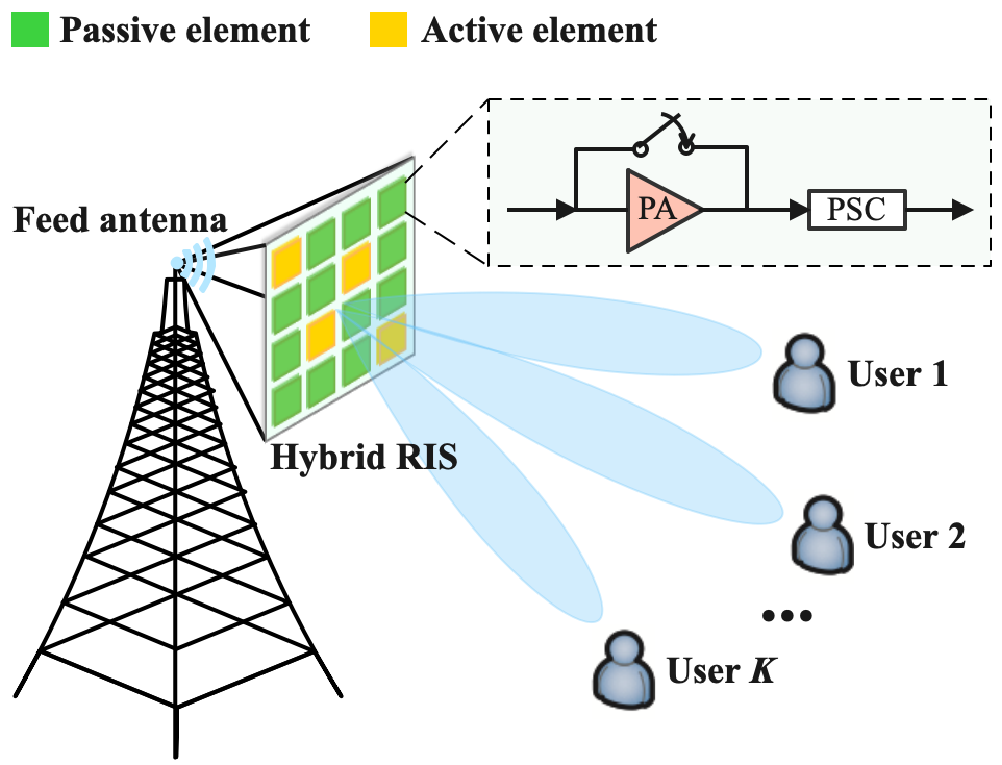}
	\caption{Illustration of the hybrid active-passive RIS transmitter enabled multi-user communication network. }
	\label{System model}
\end{figure}
As illustrated in Fig. 1, we consider a downlink multi-user communication framework, where the hybrid active-passive RIS transmitter equipped with a feed antenna serves $K$ single-antenna users simultaneously. The set of users is denoted by $\mathcal{K}=\{1,2,\cdots,K\}$. The hardware material of the RIS supports full transmission of incident electromagnetic waves, i.e. no incident signal is reflected. Suppose that the RIS is a uniform planar array (UPA)$\footnote{Remarkably, the algorithms proposed in this paper are also applicable to the case where the RIS is a uniform linear array (ULA), which can be regarded as a particular instantiation of UPA.}$ consisting of $M=M_{x} \times M_{y}$ electromagnetic elements with element spacing $\Delta_{t}$. In this paper, all channels are assumed to follow a quasi-static flat-fading channel model~\cite{mu2021capacity}. For ease of exposition, $\mathbf{h}\in\mathbb{C}^{M \times 1}$ and $\mathbf{g}_{k}\in\mathbb{C}^{M \times 1}$ are defined as the channels from the feed antenna to the hybrid RIS and the hybrid RIS to user $k\in\mathcal{K}$, respectively. To characterize the theoretical upper bound of the system performance, we assume that perfect CSI is available at the transmitter with advanced channel estimation techniques~\cite{wang2020channel,wei2021channel}.

In practical implementations, the communication between the feed antenna and the RIS occurs over a short distance, typically less than the Rayleigh distance~\cite{liu2023near}. To properly account for the unique characteristics arising from wave propagation in the near-field regime, a statistical model known as the non-uniform spherical wave-based channel model is adopted herein$\footnote{Since near-field communication leads to strong interactions and phase differences among different RIS elements, it is essential to treat each element as an independent entity~\cite{pizzo2022fourier}.}$~\cite{liu2023transmissive}.
Then, the near-field channel from the feed antenna to the RIS can be directly modeled as a line-of-sight (LoS) channel:
\begin{equation}
	\mathbf{h} = \frac{1}{\sqrt{4\pi (r_{m_{x},m_{y}})^{2}}}\left[e^{-j 2 \pi \frac{r_{1,1}}{\lambda_c}}, \ldots, e^{-j 2 \pi \frac{r_{M_{x},M_{y}}}{\lambda_c}}\right]^{H},
\end{equation}
where $\lambda_c$ denotes the carrier wavelength. $r_{m_{x},m_{y}}$ denotes the distance between the feed antenna and the $(m_{x},m_{y})$-th RIS element. $r_{m_{x}, m_{y}}=\sqrt{\bar{r}^{2}+\tilde{r}_{m_{x}, m_{y}}^2}$, where $\bar{r}$ denotes the the distance between the feed antenna and the center of the RIS, $\tilde{r}_{m_{x}, m_{y}}=\sqrt{\partial_{m_{x}}^{2} \Delta_{t}^{2}+\partial_{m_{y}}^{2} \Delta_{t}^{2}}$ denotes the distance between the $(m_{x},m_{y})$-th RIS element and the center of the RIS, $\partial_{m_{x}}=\frac{2 m_{x}-M_{x}-1}{2}$, $\partial_{m_{y}}=\frac{2 m_{y}-M_{y}-1}{2}$.

On the other hand, users are usually located outside the Rayleigh distance of the transmitter. Due to the presence of both LoS components and non-line-of-sight (NLoS) components in far-field planar wave propagation, the channel from the RIS to user $k$ can be modeled as a Rician fading channel$\footnote{Both LoS and Rician fading channels are rigorous approximations for characterizing the behavior of actual channels. Besides, it should be noted that the presented algorithms exhibit high feasibility across various channel models, i.e., the Gaussian and Rayleigh fading channels.}$:
\begin{equation}
	\mathbf{g}_{k}=\sqrt{\frac{\beta_{0}}{\left(d_{k}\right)^{\alpha}}}\left(\sqrt{\frac{\kappa}{1+\kappa}}\mathbf{g}_{k}^{\text{LoS}}+\sqrt{\frac{1}{1+\kappa}} \mathbf{g}_{k}^{\text{NLoS}}\right), \forall k\in\mathcal{K},
\end{equation}
where $\beta_{0}$  denotes the path loss at a reference distance of $1$ meter. $\alpha$ denotes the corresponding path loss exponent. $d_{k}$ denotes the distance between the RIS and user $k$. $\kappa$ denotes the Rician factor of the RIS-user link. $\mathbf{g}_{k}^{\text{LoS}}=\mathbf{a}\left(\phi_{k}, \varphi_{k}\right)$ is the deterministic LoS component and 
$\mathbf{g}_{k}^{\text{NLoS}}\sim\mathcal{CN} (\mathbf{0}, \mathbf{I}_M)$ is the NLoS component modeled as Rayleigh fading. $\mathbf{a}\left(\phi, \varphi\right)$ is the array response vector (ARV), and given by~\cite{mu2021joint}
\begin{equation}
	\begin{aligned}
		\mathbf{a}\left(\phi, \varphi\right) = & \left[1, e^{-j 2 \pi \frac{\Delta_{t} \sin \phi \cos \varphi}{\lambda_c}}, \ldots, e^{-j 2 \pi\left(M_{x}-1\right)\frac{\Delta_{t} \sin \phi \cos \varphi}{\lambda_c}}\right]^{T} \\
		&\otimes\left[1, e^{-j 2 \pi \frac{\Delta_{t} \sin \phi \sin \varphi}{\lambda_c}}, \ldots, e^{-j 2 \pi\left(M_{y}-1\right) \frac{\Delta_{t} \sin \phi \sin \varphi}{\lambda_c}}\right]^{T},
	\end{aligned}
\end{equation}
where $\phi\in\left[-\frac{\pi}{2},\frac{\pi}{2}\right]$ and $\varphi\in\left[0,2\pi\right]$ denote the elevation angle-of-arrival (AoA)/angle-of-departure (AoD) and the azimuth AoA/AoD, respectively.
\subsection{Hybrid Active-Passive RIS Transmitter Architecture}
Fig. 1 shows the the hardware architecture of the proposed hybrid RIS transmitter design. Each passive component can transmit the incident signal with a desired phase shift, while the active components are implemented by both power amplifier (PA) and phase shift control (PSC), thereby achieving simultaneous tuning over the phase and amplitude of the signal. By toggling the switches ON or OFF, the hybrid RIS elements can flexibly transition between active and passive modes to align with the practical needs of specific applications.

To mathematically characterize the operating mode of each element on the hybrid RIS, we define the transmission coefficient matrices as $\mathbf{ \Theta}={\textup{diag}}\left\{|{\beta_1}|e^{j\theta_{1}}, |{\beta_2}|e^{j\theta_{2}},\cdots, |{\beta_M}|e^{j\theta_{M}}\right\}$. $|{\beta_m}|$ and $\theta_{m}\in [0,2\pi)$, $m\in\mathcal{M}$ respectively denote the transmission amplitude and phase-shift coefficients of the $m$-th element. It is assumed that there are total $M_{\rm act}$ ($M_{\rm act} \le M$) elements operating in the active mode. The positions of active elements are predefined
in $\mathcal{A} \subset \{1, 2, \cdots, M\}$ with $|\mathcal{A}| = M_{\rm act}$. $\alpha_m \in\left\{0,1\right\}, \forall m \in \mathcal{M}$ is a RIS mode scheduling variable which is defined as:
\begin{align}
	\alpha_m\triangleq\left\{
	\begin{aligned}
		&1, \;\;\textup{Active mode at element}\;m, \;\;m\in\mathcal{A},\\
		&0, \;\;\textup{Passive mode at element}\;m,  \;\;m\notin\mathcal{A}.
	\end{aligned}
	\right.
\end{align}
Based on the above definition, matrix $\mathbf{A}=\textup{diag}\left\{\alpha_1, \alpha_2, \cdots , \alpha_M\right\} \in \mathbb{B}^{M \times M}$ is the hybrid RIS mode scheduling matrix. Furthermore, $\mathbf{\Phi} = \mathbf{A}\circ\mathbf{ \Theta}$ and $\mathbf{\Psi}= (\mathbf{I}_M-\mathbf{A})\circ\mathbf{ \Theta}$ are the diagonal active and passive coefficient matrices, respectively, where $\circ$ represents a Hadamard product.
For the passive RIS element $m\notin\mathcal{A}$, we have $|\beta_m|=1$ due to the law of energy conservation. For the active RIS element $m\in\mathcal{A}$, we have $|\beta_m|\le\rho_{\max}$, where $\rho_{\max}>1$ is the maximum power amplification gain supplied by the active load.
\subsection{Signal Model and Problem Formulation}
Let $x_k$ denote the information-bearing symbol for user $k$ sent by the transmitter, the signal received by user $k$ is expressed as
\begin{equation}
y_{k}=\left(\mathbf{g}_{k}^{H}\mathbf{\Theta} \mathbf{h}\right) \sum_{k\in\mathcal{K}} \sqrt{p_{k}} x_{k}+\mathbf{g}_{k}^{H}\mathbf{\Phi} \mathbf{n}_{r}+n_{k}, 
\end{equation}
where ${\mathbb{E}}\left[ {{{\left| {{x_k}} \right|}^2}} \right] = 1$. $n_k\sim\mathcal{CN} (0, \sigma^2)$ denotes the additive white Gaussian noise (AWGN) at user $k$ with noise power $\sigma^2_k$.
$\mathbf{n}_{r}\sim \mathcal{CN} (\mathbf{0}, \sigma_{r}^2\mathbf{I}_M)$ denotes the thermal noise generated by active RIS elements. $p_k$ denotes the transmit power allocated to user $k$ by the RF feed antenna. The received signal-to-interference-plus-noise ratio (SINR) at user $k$ is given by
\begin{equation}
	\textup{SINR}_{k}=\frac{\left| \sqrt{p_{k}}\left(\mathbf{g}_{k}^H\mathbf \Theta \mathbf{h}\right)\right|^2}{\sum_{j\in\mathcal{K}\backslash \{k\}}\left|\sqrt{p_{j}}\left(\mathbf{g}_{k}^H\mathbf \Theta \mathbf{h}\right)\right|^2+\left\|\mathbf{g}_{k}^H\mathbf \Phi\right\|^2\sigma_{r}^2+\sigma^2},
\end{equation}
$R_{k}=\log_2\left(1+\textup{SINR}_{k}\right)$ is the achievable data rate of user $k\in\mathcal{K}$.

Regarding the energy consumption model, the allocated power at the feed antenna should satisfy the following conditions,
\begin{equation}
	p_{k}>0,\quad \sum_{k\in\mathcal{K}} p_{k} \leq P_{f}^{\rm max},
\end{equation}
where $P_{f}^{\rm max}$ denotes the maximum allowable power of the feed antenna for signal transmission. Let $P_{\rm act}$ represent the amplification power of the active RIS elements, which can be expressed as~\cite{nguyen2022hybrid}
\begin{equation}
	P_{\rm act}=\sum_{k\in\mathcal{K}} p_{k}\left\|\mathbf{\Phi} \mathbf{h}\right\|^{2}+\left\|\mathbf{\Phi}\right\|^{2} \sigma_{r}^{2} \leq P_{r}^{\rm max},
\end{equation}
where $P_{r}^{\rm max}$ is the maximum allowable power at the hybrid active-passive RIS.
The total power consumption of the communication system is calculated by
\begin{equation}
	P_{\rm tot}= \frac{1}{\zeta}\left(\sum_{k\in\mathcal{K}} p_{k}+P_{\rm act}\right)+P_{\rm RF}+MP_{r}+P_{c},
\end{equation}
where $\zeta$ is the efficiency of the power amplifier, $P_{\rm RF}$ is the constant RF chain power consumption of the feed antenna, $P_{r}$ is the hardware consumption induced by each RIS element, $P_{c}$ is other circuit power consumption in the system.

In the proposed communication framework, an energy-efficient design is desired for striking a good balance between system performance and power consumption. By jointly optimizing the RIS element scheduling and beamforming coefficients, as well as the power allocation coefficients at the feed antenna, our objective is to maxmize the EE of the considered system, subject to the user's individual rate requirement and the maximum RIS amplification power constraint. Consequently, the optimization problem can be formulated as
\begin{subequations}\label{P1}
	\begin{align}
	    &\mathop{\rm{max}}\limits_{\alpha_m, \mathbf{\Theta}, p_k}\quad\eta_{\rm EE}\triangleq\frac{\overline{R}}{P_{\rm tot}}\\
		&\label{P1_C1}\quad{\rm s.t.} \;\;\,
		p_{k}>0,\quad \sum_{k\in\mathcal{K}} p_{k} \leq P_{f}^{\rm max},\\
		&\label{P1_C2}\quad\quad\quad
    	\sum_{k\in\mathcal{K}} p_{k}\left\|\mathbf{\Phi} \mathbf{h}\right\|^{2}+\left\|\mathbf{\Phi}\right\|^{2} \sigma_{r}^{2} \leq P_{r}^{\rm max},\\
		&\label{P1_C3}\quad\quad\quad
		R_{k}\ge R_k^{\rm min},\\
		&\label{P1_C4}\quad\quad\quad
		\theta_m\in[0,2\pi),  \forall m\in\mathcal{M},\\
		&\label{P1_C5}\quad\quad\quad
		|\beta_m|\le\rho_{\max},  \forall m\in\mathcal{A},\\
		&\label{P1_C6}\quad\quad\quad
		|\beta_m| =1,  \forall m\notin\mathcal{A},\\
		&\label{P1_C7}\quad\quad\quad
		\alpha_m \in\{0,1\}, \forall m \in \mathcal{M},
	\end{align}
\end{subequations}
where $\overline{R}={\sum_{k\in\mathcal{K}}R_{k}}$. Constraints \eqref{P1_C1} and \eqref{P1_C2} restrict the power allowance at the feed antenna and the hybrid RIS, respectively. Constraint \eqref{P1_C3} is imposed to guarantee that the achievable rate of user $k$ must not be lower than a rate threshold $R_k^{\rm min}$. Constraint \eqref{P1_C4} denotes the phase-shift constraint for each element on the RIS. Constraints \eqref{P1_C5} and \eqref{P1_C6} represent the amplitude coefficient constraints for the active and the passive elements, respectively. Constraint \eqref{P1_C7} indicates that each RIS element can only operate in either passive mode or active mode. 

Note that problem \eqref{P1} is a MINLP problem, which presents great difficulty in solving it directly.
Next, we will first restate problem \eqref{P1} in a nonfractional form, and then analyze the approaches to solve it. 
\subsection{Solution Analysis}
By exploiting the Dinkelbach method~\cite{fang2022energy}, problem \eqref{P1} is reformulated as
\begin{subequations}\label{P2}
	\begin{align}	
		&\mathop{\rm{max}}\limits_{\alpha_m, \mathbf{\Theta}, p_k}\quad \eta_{\rm EE}'\triangleq \overline{R}-\xi{P_{\rm tot}}\\
		&\quad\,{\rm s.t.} \;\;\,
		\textup{\eqref{P1_C1} -- \eqref{P1_C7}},
	\end{align}
\end{subequations}
where $\xi\ge 0$ is the Dinkelbach variable, which can be interpreted as a scalar weight of power consumption ${P_{\rm tot}}$. Define $\mathbf{\mathcal{Q}}\triangleq{\left\{{\alpha}_m, {\mathbf{\Theta}}, {p}_k\right\}}$. For a given $\xi$, the optimal solution of \eqref{P2} is denoted by $\mathbf{\tilde{\mathcal{Q}}}$. Assume that $\mathbf{\tilde{\mathcal{Q}}}^*$ is the optimal solution of \eqref{P1} and $\xi ^{*}$ is the maximum system EE, i.e., $\xi ^{*}=\frac{\overline{R}\left( \mathbf{\tilde{\mathcal{Q}}}^* \right)}{P_{\rm tot}\left( \mathbf{\tilde{\mathcal{Q}}}^* \right)}=\max_{\mathbf{\mathcal{Q}}}\frac{\overline{R}{\left(\mathbf{\mathcal{Q}}\right)}}{P_{\rm tot}{\left(\mathbf{\mathcal{Q}} \right)}}$. It is readily observed that solving \eqref{P1} is equivalent to finding the optimal solution $\xi ^{*}$ in problem \eqref{P2}. Define a function $F(\xi)$ as
\begin{equation}
	F(\xi)\triangleq \max_{\mathbf{\mathcal{Q}}}\left\{\overline{R}{\left(\mathbf{\mathcal{Q}}\right)}-\xi  P_{\rm tot}\left(\mathbf{\mathcal{Q}}\right)\right\},
\end{equation}
$\xi ^{*}$ can be achieved if and only if the following condition holds:
\begin{equation}
	\begin{aligned}
		F(\xi^{*})&=\max_{\mathbf{\mathcal{Q}}}\left\{\overline{R}{\left(\mathbf{\mathcal{Q}}\right)}-\xi ^{*}P_{\rm tot}\left(\mathbf{\mathcal{Q}}\right)\right\}\\	
		&=\overline{R}{\left(\mathbf{\tilde{\mathcal{Q}}}^*\right)}-\xi ^{*}P_{\rm tot}\left(\mathbf{\tilde{\mathcal{Q}}}^*\right)=0.
	\end{aligned}
\end{equation}
A two-layer algorithm framework is adopted to find $\xi ^{*}$. $\xi$ is initialized with a small value. In each iteration, $\xi$ remains constant in the inner layer and is increased in the outer layer until convergence is reached. Specifically, we can update the Dinkelbach variable $\xi$ according to $\xi^{(n+1)}=\frac{\overline{R}\left(\mathbf{\tilde{\mathcal{Q}}}^{(n)} \right)}{P_{\rm tot}\left(\mathbf{\tilde{\mathcal{Q}}}^{(n)} \right)}$, where $n$ is the iteration index. 

However, due to the fact that optimization variables are highly coupled in problem \eqref{P2}, it is challenging to obtain a globally optimal solution to this problem. Therefore, we aim to seek its locally optimal solution via iterative algorithms based on AO method. To make progress in solving binary variables, an exhaustive search-based method is developed to determine the RIS element operating mode first. Subsequently, the RIS beamforming vector and power allocation coefficients are alternately designed for a given RIS mode scheduling. On the one hand, to facilitate RIS beamforming design, semidefinite relaxation (SDR) technique is applied. Furthermore, we exploit the penalty-based method to recover the optimal rank-one solution in the SDR. On the other hand, the solution to the power allocation subproblem is achieved by successive convex approximation (SCA). Nevertheless, since the optimal $\{\alpha_m\}$ is obtained by traversing all possible index sets $\mathcal{A}$, this leads to a potentially high algorithmic complexity. In pursuit of a good complexity-optimality trade-off, a joint hybrid RIS design scheme is convinced to jointly optimize the variables $\{\alpha_m\}$ and $\{\mathbf{ \Theta}\}$ by leveraging the Big-M formulation.
The overall algorithm flowchart is shown in Fig. 2 at the top of this page. 
\begin{figure}[t]
	\centering
	\includegraphics[width=3.5in]{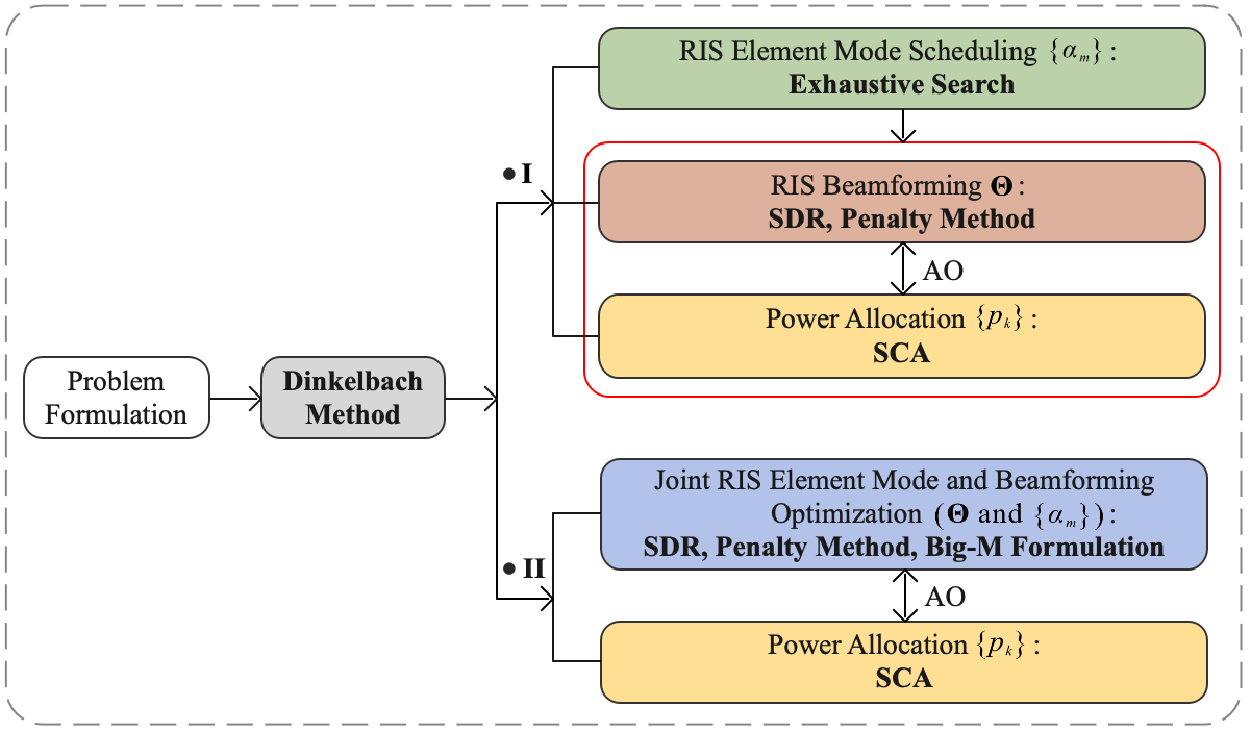}
	\caption{Algorithm flowchart.}
	\label{Proposed algorithm}
\end{figure}
\section{Alternating Optimization Based Algorithm}
In this section, we first decompose problem \eqref{P2} into three subproblems, i.e., RIS element mode scheduling optimization, RIS transmission beamforming optimization and power allocation optimization. Then, an AO-based algorithm is developed to alternately design the optimization variables.
\subsection{Element Mode Scheduling Optimization for the Hybrid RIS}
Before designing the RIS amplitude and phase-shift coefficients $\{\mathbf{ \Theta}\}$ and the feed antenna power allocation coefficients $\{p_k\}$, we need to determine the operating mode of the RIS elements, i.e., $\{\alpha_m\}$. Unfortunately, it is non-trivial to derive a closed-form solution for the optimal element mode scheduling. To circumvent this difficulty, we are going to exploit the interplay between the number of active and passive elements to shed light on the solution to problem \eqref{P2}.

Particularly, the RIS amplification power function in constraint \eqref{P1_C2} can be rewritten as:
\begin{equation}
	P_{\rm act}\overset{(a)}=\sum_{m\in\mathcal{A}}|\beta_m|^2\left(|{h}_m|^2\sum_{k\in\mathcal{K}} p_{k}+\sigma_{r}^2 \right),
\end{equation}
where ${h}_m$ is the $m$-th entry of vector $\mathbf{h}$.
Denote $\widetilde{P}\triangleq\sum_{i\in\mathcal{A},i\neq m}|\beta_i|^2\left(|{h}_m|^2\sum_{k\in\mathcal{K}} p_{k}+\sigma_{r}^2 \right)$, $\left\{|\beta_m|\right\}_ {m\in\mathcal{A}}$ can be calculated by the following equation:
\begin{equation}
	|\beta_m|=\sqrt{\frac{P_{\rm act}-\widetilde{P}}{|{h}_m|^2\sum_{k\in\mathcal{K}} p_{k}+\sigma_{r}^2}}, \forall m\in\mathcal{A}.
\end{equation}
\begin{remark}From the perspective of maximizing the system EE, $m\in\mathcal{A}$ is available if and only if $|\beta_m|>1$ holds. According to Eq. (15), $\left\{|\beta_m|\right\}_ {m\in\mathcal{A}}$ is strictly limited by the amlification power budget $P_{\rm act}$. Specifically, by fixing $P_{\rm act}$ with $P_{\rm act}\le P_{r}^{\rm max}$, for given power allocation coefficients $\left\{p_{k}\right\}$, a larger number of active RIS elements (i.e., $M_{\rm act}$) will lead to a smaller $|\beta_m|$. Therefore, increasing $M_{\rm act}$ may compress the amplification of the transmitted signal amplitude, and even result in $|\beta_m|\le1$, which attenuates the signals and degrades the system performance. This conclusion will be further demonstrated numerically.
\end{remark}
On the other hand, we note that thermal noise caused by active elements cannot be ignored and should be taken into account when calculating the system data rate. An RIS with a higher quantity of active elements also requires an increased power consumption~\cite{long2021active}. Hence, some of the elements on the hybrid RIS need to be switched to the passive mode to guarantee optimal EE performance at a limited amplification power threshold. Based on the above observations, a special case is considered to help us find the maximum $M_{\rm act}$. For any active element $m\in\mathcal{A}$, by setting $|\beta_m|=1$, problem \eqref{P2} is feasible when the following condition holds:
\begin{equation}
	\sum_{m\in\mathcal{A}}|{h}_m|^2\le\frac{P_{r}^{\rm max}-\sigma_{r}^2}{\sum_{k\in\mathcal{K}} p_{k}}\triangleq \psi\left(\mathbf{h}\right).
\end{equation}
For the near-field channel $\mathbf{h}\in\mathbb{C}^{M \times 1}$, we first rank $|{h}_m|, \forall m\in\mathcal{A}$ in an increasing order as
\begin{equation}
	|[{h}_m]_{j_1}|\le|[{h}_m]_{j_2}|\le\cdots\le|[{h}_m]_{j_M}|,
\end{equation}
where $j_i$ is the index corresponding to the rank ordering. It is assumed that there exists
\begin{align}
	&\sum_{i=1}^{L}|[{h}_m]_{j_i}|^2\le\psi\left(\mathbf{h}\right),\\
	&\sum_{i=1}^{L+1}|[{h}_m]_{j_i}|^2>\psi\left(\mathbf{h}\right),
\end{align}
based on (18) and (19), we are able to obtain the upper bound of $M_{\rm act}$, which is given by 
\begin{equation}
	[M_{\rm act}]^{ub}=\min\left\{L, M\right\}.
\end{equation}

For an $M$ element hybrid RIS with $M_{\rm act}$ active elements, given the upper bound of $M_{\rm act}$, i.e., $M_{\rm act}\le[M_{\rm act}]^{ub}$, there are $P_{\mathcal{A}}=\sum_{i=1}^{[M_{\rm act}]^{ub}}\binom{M}{i}$ different candidate sets $\mathcal{A}$. We exhaustively search through all possible index sets $\mathcal{A}$ to find the optimal RIS mode scheduling scheme. For given $\{\alpha_m\}$, we could alternately solve the following two subproblems by applying the AO method. Ultimately, the scheme can obtain the highest objective value will be selected as the optimal solution.
\subsection{Transmission Beamforming Optimization at the Hybrid RIS}
Given $\{\alpha_m\}$ and $\{p_k\}$, the problem for optimizing RIS transmission beamforing coefficients $\{\mathbf{ \Theta}\}$ becomes
\begin{subequations}\label{P3}
	\begin{align}	
		&\mathop{\rm{max}}\limits_{\mathbf{ \Theta}}\quad\eta_{\rm EE}'\\
		&\;\;{\rm s.t.} \;\;	
		\label{P3_C1}\textup{SINR}_{k}\ge\gamma_k^{\rm min},\\
		& \quad\quad\;\,	\textup{\eqref{P1_C2}, \eqref{P1_C4} -- \eqref{P1_C6}},
	\end{align}
\end{subequations}
where constraint \eqref{P3_C1} is a variant of constraint \eqref{P1_C3}, $\gamma_k^{\rm min}=2^{R_k^{\rm min}}-1$ is the required minimum SINR for user $k$. The non-convexity of problem \eqref{P3} lies in the non-concave objective function and the non-convex constraints \eqref{P1_C2} and \eqref{P3_C1}. 

Since matrices $\mathbf{A}$ and $\mathbf{ \Theta}$ are two diagonal matrices, we can rewrite their Hadamard product in a normal product form, i.e., $\mathbf{\Phi} = \mathbf{A}\mathbf{ \Theta}$. Therefore, constraint \eqref{P1_C2} can be rewritten as
\begin{equation}
	\sum_{k\in\mathcal{K}} p_{k}\left\|\mathbf{A}\mathbf{ \Theta} \mathbf{h}\right\|^{2}+\left\|\mathbf{A}\mathbf{ \Theta}\right\|^{2} \sigma_{r}^{2} \leq P_{r}^{\rm max}.
\end{equation}
To tackle the non-convexity of the quadratic term in constraint (22), define the transmission coefficient vector at the hybrid RIS as $\mathbf{u}\triangleq[|{\beta_1}|e^{j\theta_{1}}, |{\beta_2}|e^{j\theta_{2}},\cdots, |{\beta_M}|e^{j\theta_{M}}]^H\in\mathbb{C}^{M \times 1}$, the left-hand side term of constraint (22) can be further expressed as $P_{\rm act}=\sum_{k \in \mathcal{K}}p_k\left\|\mathbf{u}^H\mathbf{H}{\mathbf{A}}\right\|^{2}+\sigma_{t}^{2}\|\mathbf{u}^H{\mathbf{A}}\|^{2}$, where $\mathbf{H}\triangleq\textup{diag}\left({\mathbf{h}}^H\right)\in\mathbb{C}^{M \times M}$.
By lifting ${\mathbf{A}}$ into matrix $\widetilde{\mathbf{A}}={\mathbf{A}}{\mathbf{A}}^H\in\mathbb{C}^{M \times M}$, then the hybrid RIS amplification power constraint is approximated by the convex constraint as follows:
\begin{equation}
	\sum_{k \in \mathcal{K}}{p_{k}\textup{Tr}\left(\mathbf{U} {\mathbf{H}} \widetilde{\mathbf{A}} {\mathbf{H}}^{H}\right)}+{\textup{Tr}\left( \widetilde{\mathbf{A}}\mathbf{U}\right)}\sigma_{r}^{2}\le P_{r}^{\rm max},
\end{equation}
where $\mathbf{U}={\mathbf{u}}{\mathbf{u}}^H\in\mathbb{C}^{M \times M}$, satisfying $\mathbf U\succeq 0$ and $\textup{Rank}(\mathbf U)=1$. Denote $\mathbf{G}_k\triangleq\textup{diag}\left(\mathbf{g}_{k}^H\right)\in\mathbb{C}^{M \times M}$, the SINR term in (6) is given by
\begin{equation}
	\begin{aligned}
		&\textup{SINR}_{k}=\frac{{p_{k}}\left|\mathbf{u}^H\mathbf{G}_k\mathbf{h}\right|^2}{\sum_{j\in\mathcal{K}\backslash \{k\}}{p_{j}}\left|\mathbf{u}^H\mathbf{G}_k\mathbf{h}\right|^2+\left\|\mathbf{u}^H\mathbf{G}_k\mathbf{A}\right\|^2\sigma_{r}^2+\sigma^2}\\
		&=\frac{{p_{k}}\textup{Tr}\left(\mathbf{U} {\mathbf{G}_k} \widetilde{\mathbf{H}} {\mathbf{G}_k^{H}}\right)}{\sum_{j\in\mathcal{K}\backslash \{k\}}{p_{j}}\!\textup{Tr}\left(\mathbf{U} {\mathbf{G}_k} \widetilde{\mathbf{H}} {\mathbf{G}_k^{H}}\right)\!+\!\textup{Tr}\left(\mathbf{U} {\mathbf{G}_k} \widetilde{\mathbf{A}} {\mathbf{G}_k^{H}}\right)\!\sigma_{r}^2\!+\!\sigma^2},
	\end{aligned}
\end{equation}
where $\widetilde{\mathbf{H}}={\mathbf{h}}{\mathbf{h}}^H\in\mathbb{C}^{M \times M}$. The SINR term is now converted into a fractional function with a linear numerator over a linear denominator.
Therefore, constraint \eqref{P3_C1} can be rewritten as
\vspace{0.3cm}
\begin{equation}
	\begin{aligned}
		&\gamma_k^{\rm min}{\left(\!\sum_{j\in\mathcal{K}\backslash \{k\}}\!\!{p_{j}}\textup{Tr}\left(\mathbf{U} {\mathbf{G}_k} \widetilde{\mathbf{H}} {\mathbf{G}_k^{H}}\right)\!+\!\textup{Tr}\left(\mathbf{U} {\mathbf{G}_k} \widetilde{\mathbf{A}} {\mathbf{G}_k^{H}}\right)\sigma_{r}^2\!+\!\sigma^2\!\right)}\\
		&\le{{p_{k}}\textup{Tr}\left(\mathbf{U} {\mathbf{G}_k} \widetilde{\mathbf{H}} {\mathbf{G}_k^{H}}\right)}, \forall k \in\mathcal{K}.
	\end{aligned}
\end{equation}
Introduce the following slack variables $\{X_{k}\}$ and $\{Y_{k}\}$: 
\begin{gather}
	\frac{1}{X_{k}}\le{p_{k}}\textup{Tr}\left(\mathbf{U} {\mathbf{G}_k} \widetilde{\mathbf{H}} {\mathbf{G}_k^{H}}\right),\\
	Y_{k}\ge\!\sum_{j\in\mathcal{K}\backslash \{k\}}\!\!{p_{j}}\textup{Tr}\left(\mathbf{U} {\mathbf{G}_k} \widetilde{\mathbf{H}} {\mathbf{G}_k^{H}}\right)\!+\!\textup{Tr}\left(\mathbf{U} {\mathbf{G}_k} \widetilde{\mathbf{A}} {\mathbf{G}_k^{H}}\right)\sigma_{r}^2\!+\!\sigma^2,
\end{gather}
for any given points $\{{X_{k}^{(n)}},{Y_{k}^{(n)}}\}$, based on the first-order Taylor expansion~\cite{boyd2004convex}, the concave lower bound of the achievable rate at user $k$ can be obtained as
\begin{equation}
	\begin{aligned}
		R_{k}\ge&\log_2\left(1+\frac{1}{{X_{k}^{(n)}}{Y_{k}^{(n)}}}\right)-\frac{\left(\log_2e\right)\left({X_{k}}-{X_{k}^{(n)}}\right)}{{X_{k}^{(n)}}+{X_{k}^{(n)}}^2{Y_{k}^{(n)}}}\\&-\frac{\left(\log_2e\right)\left({Y_{k}}-{Y_{k}^{(n)}}\right)}{{Y_{k}^{(n)}}+{Y_{k}^{(n)}}^2{X_{k}^{(n)}}}\triangleq\left[\Pi_{k}\right]^{lb}, \forall k \in\mathcal{K}.
	\end{aligned}
\end{equation}

Then, the RIS beamforming optimization problem can be approximated as
\begin{subequations}\label{P4}
	\begin{align}
		&\mathop{\rm{max}}\limits_{\mathbf{U}, X_{k}, Y_{k}}\quad 
		{\sum_{k\in\mathcal{K}}\left[\Pi_{k}\right]^{lb}}-\xi{P_{\rm tot}}\\
		&\label{P4_C1}\;\;{\rm s.t.} \;\;
		\mathbf U \succeq 0,\\	
		&\label{P4_C2}\qquad\;\,
		\text{Rank}\left(\mathbf{U}\right)=1,\\	
		&\label{P4_C3}\qquad\;\,
		\left|\mathbf{U}\right|_{mm}\le\rho_{\max},  \forall m\in\mathcal{A},\\
		&\label{P4_C4}\qquad\;\,
		\left|\mathbf{U}\right|_{mm}=1,  \forall m\notin\mathcal{A},\\
		&\qquad\;\,
		\textup{(23), (25) -- (27)}.
	\end{align}
\end{subequations}
Observed that the main challenge in solving problem \eqref{P4} stems from coping with the non-convex rank-one constraint \eqref{P4_C2}. Using the SDR technique, we can always obtain an optimal solution to the relaxed version of the above problem, i.e., by dropping constraint \eqref{P4_C2}.

In the following, penalty method is employed to ensure the rank-one property of $\mathbf{U}$. To elaborate it, we first provide an equivalent representation for constraint \eqref{P4_C2} as $\left\|\mathbf{U}\right\|_*-\left\|\mathbf{U}\right\|_2=0$~\cite{mu2021simultaneously}, which can be appended to (29a) as a penalty function term. Here, $\left\|\mathbf{U}\right\|_*\triangleq\sum_i\sigma_i\left(\mathbf{U}\right)$ and $\left\|\mathbf{U}\right\|_2\triangleq \sigma_1\left(\mathbf{U}\right)$ denote the nuclear norm and spectral norm, respectively, and $\sigma_i\left(\mathbf{U}\right)$ is the $i$-th largest singular value of matrix $\mathbf{U}$. Now, the objective function of the optimization problem is transformed into
\begin{equation}
	\eta_{\rm EE}'\left({\mathbf{U}, X_{k}, Y_{k}}\right)\triangleq \left({\sum_{k\in\mathcal{K}}\left[\Pi_{k}\right]^{lb}}-\xi{P_{\rm tot}}\right)-\tau\left( \left\|\mathbf{U}\right\|_*-\left\|\mathbf{U}\right\|_2\right),
\end{equation}
where $\tau$ is the penalty factor. In order to guarantee the concavity of the objective function, SCA technique is applied to construct an upper bound on the penalty function term as follows:
\begin{equation}
	\begin{aligned}
		\left\|\mathbf{U}\right\|_*-\left\|\mathbf{U}\right\|_2\le\left\|\mathbf{U}\right\|_*-\overline{\mathbf U}^{(n)}\triangleq {\chi\left(\mathbf{U},\mathbf{U}^{(n)}\right)},
	\end{aligned}
\end{equation}
where $\overline{\mathbf U}^{(n)}\triangleq\left\|\mathbf U^{(n)}\right\|_2+\textup{Tr}\left(\delta_{\rm max}\left(\mathbf U^{(n)}\right)\delta_{\rm max}\left(\mathbf U^{(n)}\right)^H\left(\mathbf{U}-\mathbf U^{(n)}\right)\right)$.
$\mathbf U^{(n)}$ denotes the given local point in the $l$-th iteration, $\delta_{\rm max}\left(\mathbf U^{(n)}\right)$ denotes the eigenvector corresponding to the largest eigenvalue of $\mathbf U^{(n)}$. 

As a result, by replacing the non-convex penalty function term with its convex upper bound, we obtain a standard convex optimization problem as follows:
\begin{subequations}\label{P5}
	\begin{align}
		&\mathop{\rm{max}}\limits_{\mathbf{U}, X_{k}, Y_{k}}\quad 
		\left({\sum_{k\in\mathcal{K}}\left[\Pi_{k}\right]^{lb}}-\xi{P_{\rm tot}}\right)-\tau{\chi\left(\mathbf{U},\mathbf{U}^{(n)}\right)}\\
		&\;\;{\rm s.t.} \;\;
		\textup{(23), (25) -- (27), \eqref{P4_C1}, \eqref{P4_C3}, \eqref{P4_C4}}.
	\end{align}
\end{subequations}
This problem can be efficiently solved by adopting existing convex problem solvers such as CVX.  
Denote $\mathbf{U}^*$ as the optimal solution to problem \eqref{P5}, the transmit beamforming vector can be obtained via Cholesky decomposition, i.e., $\mathbf{U}^*=\mathbf{u}\mathbf{u}^H$. Then, we can recover $\mathbf{\Theta}^*=\textup{diag}\left(\mathbf{u}\right)$.
\subsection{Power Allocation Optimization at the Feed Antenna}
Lastly, given $\{\alpha_m\}$ and $\{\mathbf{ \Theta}\}$, problem \eqref{P2} is reduced to finding the optimal power allocation coefficients $\{p_k\}$ at the feed antenna, which yields the following problem:
\begin{subequations}\label{P6}
	\begin{align}	
		&\mathop{\rm{max}}\limits_{p_k}\quad\eta_{\rm EE}'\\
		&\;\;{\rm s.t.} \;\;	
		\textup{\eqref{P1_C1}, \eqref{P1_C2}, \eqref{P3_C1}}.
	\end{align}
\end{subequations}
The objective function in problem \eqref{P6} is non-concave and the constraint \eqref{P3_C1} is non-convex, which makes the power allocation very challenging. To tackle this obstacle, let us denote $\Upsilon_k\triangleq\left\|\mathbf{g}_{k}^H\mathbf \Phi\right\|^2\sigma_{r}^2+\sigma^2$, we first transform constraint \eqref{P3_C1} as
\begin{equation}
	\gamma_k^{\rm min}{\left(\sum_{j\in\mathcal{K}\backslash \{k\}}{p_{j}}\left|\mathbf{g}_{k}^H\mathbf \Theta \mathbf{h}\right|^2+\Upsilon_k\right)}\le{p_{k}\!\left|\mathbf{g}_{k}^H\mathbf \Theta \mathbf{h}\right|^2}.
\end{equation}
The achievable rate for signal decoding at user $k$ can be equivalently rewritten as
\begin{equation}
	\begin{aligned}
		R_{k}&=\log_2\left(1+\frac{p_{k}\left|\mathbf{g}_{k}^H\mathbf \Theta \mathbf{h}\right|^2}{\sum_{j\in\mathcal{K}\backslash \{k\}}p_{j}\left|\mathbf{g}_{k}^H\mathbf \Theta \mathbf{h}\right|^2+\Upsilon_k}\right),\\
		&=\overline f_1\left(p_k\right)-\overline f_2\left(p_k\right),
	\end{aligned}
\end{equation}
where
\begin{gather}
	\overline f_1\left(p_k\right)=\log_2\left({\sum_{k\in\mathcal{K}}p_{k}\left|\mathbf{g}_{k}^H\mathbf \Theta \mathbf{h}\right|^2+\Upsilon_k}\right),\\
	\overline f_2\left(p_k\right)=\log_2\left(  {\sum_{j\in\mathcal{K}\backslash\{k\}}p_{j}\left|\mathbf{g}_{k}^H\mathbf \Theta \mathbf{h}\right|^2+\Upsilon_k} \right).
\end{gather}
Therefore, the objective function of the optimization problem is given by
\begin{equation}
	\eta_{\rm EE}'\left(p_k\right)\triangleq \sum_{k\in\mathcal{K}}\left(\overline f_1\left(p_k\right)-\overline f_2\left(p_k\right)\right)-\xi{P_{\rm tot}},
\end{equation}
where $P_{\rm tot}$ in (38) is a linear function with respect to $\left\{p_k\right\}$. It should be noted that the data rate of the system is reconstructed in terms of the difference of two logarithmic functions, its global concave underestimator can be obtained by resorting to the SCA method. Define $Q_k\triangleq\sum_{j\in\mathcal{K}\backslash \{k\}}p_j$, for any feasible point $Q_{k}^{(n)}$, we have
\begin{equation}
	\begin{aligned}
		R_{k}\ge&\overline f_1\left(p_k\right)-\log_2\left(  {Q_{k}^{(n)}\left|\mathbf{g}_{k}^H\mathbf \Theta \mathbf{h}\right|^2+\Upsilon_k} \right)\\
		&-\frac{\left(Q_{k}-Q_{k}^{(n)} \right)\left|\mathbf{g}_{k}^H\mathbf \Theta \mathbf{h}\right|^2\log_2e}{  {Q_{k}^{(n)}\left|\mathbf{g}_{k}^H\mathbf \Theta \mathbf{h}\right|^2+\Upsilon_k}}
		\triangleq [\Gamma_k]^{lb}, \forall k\in\mathcal{K}.
	\end{aligned}
\end{equation}

Till now, problem \eqref{P6} for transmit power allocation can be reformulated as
\begin{subequations}\label{P7}
	\begin{align}	
		&\mathop{\rm{max}}\limits_{p_k}\quad {\sum_{k\in\mathcal{K}}\left[\Gamma_{k}\right]^{lb}}-\xi{P_{\rm tot}}\\
		&\;\;{\rm s.t.} \quad
		\textup{\eqref{P1_C1}, \eqref{P1_C2}, (34)}.
	\end{align}
\end{subequations}
The above problem a standard convex optimization problem which can be solved directly with convex optimization software, such as CVX. 
\subsection{Proposed Algorithm and Computational Complexity}
To start with, we propose an exhaustive search-based method to determine the RIS element selection mode. Next, two subproblems are solved for a given element mode scheduling scheme by adopting the AO method. More particularly, the RIS beamforming coefficients $\{\mathbf{ \Theta}\}$ and the feed antenna power allocation coefficients $\{p_k\}$ are alternately optimized by solving problem \eqref{P5} and \eqref{P7}. The details of the proposed algorithm are summarized in \textbf{Algorithm 1}. As shown, in the $n_2$-th iteration, if the increment of $\xi$ falls below a predefined threshold $\epsilon_2$, the iteration process terminates and returns the optimal resource allocation scheme and system EE. Otherwise, we update $\xi^{(n_2+1)}=\frac{\overline{R}{\left(\mathbf{\Theta}^{(n_1)}, p_k^{(n_1)}\right)}}{P_{\rm tot}\left(\mathbf{\Theta}^{(n_1)}, p_k^{(n_1)}\right)}$ and continue the iteration process. As a result, the proposed algorithm is ensured to reach convergence. However, the assertion of attaining global optimality lacks empirical support for the following reasons: 1) Problem \eqref{P2} is not jointly convex with respect to $\mathbf{\Theta}$ and $\{p_k\}$; and 2) The optimization results obtained are based on both Dinkelbach and AO methods, which are generally not guaranteed to yield globally optimal solutions.

In the following, the computational complexity of \textbf{Algorithm 1} for soving problem \eqref{P2} is analyzed. The complexity of exhaustively searching all possible index sets $\mathcal{A}$ is ${{\mathcal{O}}}\left(P_{\mathcal{A}}\right)$. To optimize the RIS beamforming coefficients, the complexity for solving problem \eqref{P5} is $\mathcal{O}\left(\max \left(M,3K+1\right)^{4}\sqrt{M}\log _{2}\frac{1}{\epsilon}\right)$, where $\epsilon>0$ is the solution accuracy. According to~\cite{boyd2004convex}, the complexity of the power allocation design is $\mathcal{O}\left(K^{3.5}\right)$, supposing that the interior-point method is applied. In consequence, the total complexity of \textbf{Algorithm 1} is ${{\mathcal{O}}}\left(P_{\mathcal{A}}\left(I_{\rm out}I_{\rm inn}\left(\max \left(M, 3K+1\right)^{4} \sqrt{M}\log _{2}\frac{1}{\epsilon}+K^{3.5}\right)\right)\right)$, where $I_{\rm inn}$ and $I_{\rm out}$ respectively denote the number of iterations in the inner layer and outer layer required by \textbf{Algorithm 1}.

\begin{algorithm}[!t]\label{method1}
	\caption{Proposed Algorithm for Solving Problem \eqref{P2}}
	\begin{algorithmic}[1]
		\STATE Calculate the upper bound of $M_{\rm act}$ based on Eq. (20).\\
		\STATE {\bf Exhaustive searching:}\\
		\STATE \quad{\bf Initialize} $\mathcal{A}$, feasible points $\left\{\mathbf{\Theta}^{(0)}, p_k^{(0)}\right\}$, energy efficiency $\xi^{(0)}=0$, and the maximum error tolerance $\epsilon_1$ and $\epsilon_2$.\\
		\quad Set iteration index number $n_1=n_2=0$.
		\STATE \quad{\bf repeat: outer loop}
		\STATE \quad\quad {\bf repeat: inner loop}
		\STATE \quad\quad\quad Obtain $\mathbf{\Theta}^{(n_1+1)}$ by solving problem \eqref{P5} with given $\left\{\alpha_m\right\}$, $\xi^{(n_2)}$ and $\left\{p_k^{(n_1)}\right\}$.\\
		\STATE \quad\quad\quad Obtain $\left\{p_k^{(n_1+1)}\right\}$ by solving problem \eqref{P7} with given $\left\{\alpha_m\right\}$, $\xi^{(n_2)}$ and  $\mathbf{\Theta}^{(n_1+1)}$.\\
		\STATE \quad\quad\quad $n_1\leftarrow n_1+1$.
		\STATE \quad\quad {\bf until} the fractional increase of the objective value is below $\epsilon_1$ or the maximum number of inner iterations is reached.
		\STATE \quad\quad Update $\xi^{(n_2+1)}=\frac{\overline{R}{\left(\mathbf{\Theta}^{(n_1)}, p_k^{(n_1)}\right)}}{P_{\rm tot}\left(\mathbf{\Theta}^{(n_1)}, p_k^{(n_1)}\right)}$.	
		\STATE \quad\quad $n_2\leftarrow n_2+1$, and $n_1\leftarrow 0$.
		\STATE \quad{\bf until} $\left|\xi^{(n_2)}-\xi^{(n_2-1)}\right|\le\epsilon_2$ or the maximum number of outer iterations is reached.\\
		\STATE {\bf Output} the optimal solutions and maximum energy efficiency $\xi$.
	\end{algorithmic}
\end{algorithm}
\section{Low-Complexity Algorithm}
In Section III, we propose to leverage the AO method for optimizing different variables (i.e., $\{\alpha_m\}$, $\{\mathbf{ \Theta}\}$ and $\{p_k\}$). To find the optimal index set $\mathcal{A}$, an exhaustive search-based scheme is developed. It requires ${{\mathcal{O}}}\left(P_{\mathcal{A}}\right)$ computational complexity for searching, which is unacceptable even for moderate values of $M$. In this section, we present a forward-looking vision outlining the optimization for joint hybrid RIS design. Particularly, given the power allocation coefficients $\{p_{k}\}$, the element mode scheduling $\{\alpha_m\}$ and the transmission beamforming coefficients $\{\mathbf{ \Theta}\}$ of the hybrid RIS can be jointly optimized with low-complexity.
\subsection{Joint RIS Element Mode and Beamforming Optimization}
Specifically, to achieve low-complexity design for the hybrid RIS, problem \eqref{P2} is simplified as
\begin{subequations}\label{P8}
	\begin{align}
		&\mathop{\rm{max}}\limits_{\alpha_m, \mathbf{ \Theta}}\quad \eta_{\rm EE}'\\
		&\;\;{\rm s.t.} \;\;	
		\textup{\eqref{P1_C2}, \eqref{P1_C4} -- \eqref{P1_C7}, \eqref{P3_C1}}.
	\end{align}
\end{subequations}
The non-convexity of problem \eqref{P8} lies in the non-concave objective function and the non-convex constraints \eqref{P1_C2}, \eqref{P1_C7} and \eqref{P3_C1}. 

As can be seen, the optimization variables $\mathbf{A}$ and $\mathbf{ \Theta}$ are highly coupled together. Let us define $\mathbf{u}\triangleq[u_1, u_2,\cdots, u_M]^H\in\mathbb{C}^{M \times 1}$.
To tackle the coupling between $\mathbf{A}$ and $\mathbf{ \Theta}$, a slack optimization variable $\mathbf{z}\triangleq[z_1, z_2,\cdots, z_M]^H\in\mathbb{C}^{M \times 1}$ is introduced, which satisfies $\textup{diag}\left( \mathbf{z}^H\right)=\mathbf{A}\mathbf{ \Theta}$.
According to the Big-M formulation~\cite{cai2020joint}, we present a set of equivalent constraints for $\textup{diag}\left( \mathbf{z}^H\right)=\mathbf{A}\mathbf{ \Theta}$ as follows:
\begin{align}
	&-\alpha_m\tilde{C}\le z_m,\\
	&z_m\le \alpha_m\tilde{C},\\
	&u_m-\left(1-\alpha_m\right)\tilde{C}\le z_m,\\
	&z_m\le u_m+\left(1-\alpha_m\right)\tilde{C},
\end{align}
where $u_m$ and $z_m$ denote the $m$-th entry of vector $\mathbf{u}$ and $\mathbf{z}$, respectively. $z_m$ is forced to zero by constraints (42) and (43) when $\alpha_m = 0$; $z_m$ reaches the same value as $u_m$ by (44) and (45) when $\alpha_m = 1$. $\tilde{C}$ is an arbitrarily large constant. For all elements on the hybrid RIS, we have the following constraint as
\begin{equation}
	1\le |u_m|\le \rho_{\max}, \forall m.
\end{equation}
Moreover, the unit-modulus constraint should be imposed to strictly limit the amplitude of the passive elements (i.e., when $\alpha_m = 1$). Accordingly, the constraint is expressed as
\begin{equation}
	|[\left(\mathbf{I}-\mathbf{A}\right)\mathbf{ \Theta}]_{mm}|=	[\mathbf{I}-\mathbf{A}]_{mm}.
\end{equation}
To support efficient algorithm design, the same method mentioned in~\cite{wu2022wpcn} is adopted to relax constraint (47) as
\begin{equation}
	|[\textup{diag}\left( \mathbf{u}^H\right)-\textup{diag}\left( \mathbf{z}^H\right)]_{mm}|\le[\mathbf{I}-\mathbf{A}]_{mm}.
\end{equation}

After straightforward algebraic multiplications, the SINR term in (6) can be calculated by 
	\begin{equation}
		\begin{aligned} \label{eqn2}
			&\textup{SINR}_{k}\\
			&=\frac{{p_{k}}\left| \mathbf{g}_{k}^H\textup{diag}\left( \mathbf{u}^H\right) \mathbf{h}\right|^2}{\sum_{j\in\mathcal{K}\backslash \{k\}}{p_{j}}\left| \mathbf{g}_{k}^H\textup{diag}\left( \mathbf{u}^H\right) \mathbf{h}\right|^2+\left\| \mathbf{g}_{k}^H\textup{diag}\left( \mathbf{z}^H\right)\right\|\sigma_{r}^2+\sigma^2},\\
			&=\frac{{p_{k}}\left|\mathbf{u}^H\mathbf{G}_k\mathbf{h}\right|^2}{\sum_{j\in\mathcal{K}\backslash \{k\}}{p_{j}}\left|\mathbf{u}^H\mathbf{G}_k\mathbf{h}\right|^2+\left\|\mathbf{z}^H\mathbf{G}_k\right\|^2\sigma_{r}^2+\sigma^2}.
		\end{aligned}
	\end{equation}
Now, constraint \eqref{P3_C1} is equivalent to
\begin{equation}
	\begin{aligned}
		&\gamma_k^{\rm min}{\left({\sum_{j\in\mathcal{K}\backslash \{k\}}{p_{j}}\left|\mathbf{u}^H\mathbf{G}_k\mathbf{h}\right|^2+\left\|\mathbf{z}^H\mathbf{G}_k\right\|^2\sigma_{r}^2+\sigma^2}\right)}\\
		&\le{{p_{k}}\left|\mathbf{u}^H\mathbf{G}_k\mathbf{h}\right|^2}, \forall k \in\mathcal{K}.
	\end{aligned}
\end{equation}
For the non-convex constraint (50), the right-hand side (RHS) is a convex function with respect to $\mathbf{u}$, its concave lower bound with the first-order Taylor expansion is obtained as
\begin{equation}
	\begin{aligned}
		\left|\mathbf{u}^H\mathbf{G}_k\mathbf{h}\right|^2&\ge\left|\mathbf{u}^{(n)^H}\mathbf{G}_k\mathbf{h}\right|^2\\
		&+2\textup{Re}\left(\left( \mathbf{u}^{(n)^H}\mathbf{G}_k\mathbf{h} \mathbf{h}^H\mathbf{G}_k^H  \right)\left(\mathbf{u}-\mathbf{u}^{(n)} \right)\right)\\
		&\triangleq \mathcal{I}_k, \forall k \in\mathcal{K},
	\end{aligned}
\end{equation}
where $\mathbf{u}^{(n)}$ denotes the local point generated in the $l$-th iteration of the SCA method. 

Note that the optimization variables ${\alpha_m}$ for mode scheduling are binary and thus problem \eqref{P8} involves integer constraints, which makes the RIS configuration more challenging. To circumvent this difficulty, we first relax the binary variables in \eqref{P1_C7} into continuous variables, which yieds
\begin{equation}
	0\le\alpha_m\le 1, \forall m\in\mathcal{M}.
\end{equation}
More explicitly, the binary constraint \eqref{P1_C7} can be further rewritten as the following equality constraint:
\begin{equation}
	\alpha_m\left(1-\alpha_m\right)=0, \forall m\in\mathcal{M}.
\end{equation}
We can readily deduce that the value of $\alpha_m$ satisfying constraint (53) must be either $1$ or $0$, otherwise $ \alpha_m\left(1-\alpha_m\right)>0$ always holds, which confirms the equivalence of the two constraints \eqref{P1_C7} and (53). Based on this observation, the penalty method is leveraged to solve the non-convex binary constraint. In particular, we transform constraint (53) into a penalty term appended to (41a), and recast problem \eqref{P8} as
\begin{subequations}\label{P9}
	\begin{align}
		&\mathop{\rm{max}}\limits_{\alpha_m, \mathbf{u}, \mathbf{z}}\quad \left(\overline{R}-\xi{P_{\rm tot}}\right)-\mu \sum_{m\in\mathcal{M}}\left[ \alpha_m\left(1-\alpha_m\right)\right]\\
		&\label{P9_C1}\;\;{\rm s.t.}\quad
		\gamma_k^{\rm min}{\left({\sum_{j\in\mathcal{K}\backslash \{k\}}{p_{j}}\left|\mathbf{u}^H\mathbf{G}_k\mathbf{h}\right|^2+\left\|\mathbf{z}^H\mathbf{G}_k\right\|^2\sigma_{r}^2+\sigma^2}\right)}\notag\\
		&\qquad\;\;\,\le{{p_{k}}\mathcal{I}_k}, \forall k \in\mathcal{K}.\\
		&\label{P9_C2}\quad\quad\;\;\,
		\sum_{k\in\mathcal{K}} p_{k}\left\|\mathbf{z}^H\mathbf{H}\right\|^{2}+\left\|\mathbf{z}\right\|^{2} \sigma_{r}^{2} \leq P_{r}^{\rm max},\\
		&\label{P9_C3}\quad\quad\;\;\,
		\textup{(42) -- (46), (48), (52)},
	\end{align}
\end{subequations}
where $\mu>0$ is the penalty factor. We note tht the non-convexity of problem \eqref{P9} lies in the non-concave objective function (54a). 

In the previous section, we have already demonstrated how to deal with the non-convex rate function. For the penalty term in the objective function, with given point $\alpha_m^{(n)}$, we can construct its upper bound as
\begin{equation}
	\begin{aligned}
		\alpha_m\left(1-\alpha_m\right)&\le\alpha_m-{\alpha_m^{(n)}}^2-2{\alpha_m^{(n)}}\left({\alpha_m}-{\alpha_m^{(n)}}\right)\\
		&\triangleq\Omega\left(\alpha_m, \alpha_m^{(n)}\right), \forall m\in\mathcal{M}.
	\end{aligned}
\end{equation}
Now, problem \eqref{P9} can be conservatively reformulated as
\begin{subequations}\label{P10}
	\begin{align}	
		&\mathop{\rm{max}}\limits_{\alpha_m, \mathbf{u}, \mathbf{z}, X_k, Y_k}\quad \left({\sum_{k\in\mathcal{K}}R_{k}}-\xi{P_{\rm tot}}\right)-\mu \sum_{m\in\mathcal{M}}\Omega\left(\alpha_m, \alpha_m^{(n)}\right)\\
		&\label{P10_C1}\qquad{\rm s.t.}\quad R_{k}\le\left[\Pi_{k}\right]^{lb},\\
		&\label{P10_C2}\qquad\qquad\;	\frac{1}{X_{k}}\le{p_{k}}\mathcal{I}_k,\\
		&\label{P10_C3}\qquad\qquad\;	Y_{k}\ge{\sum_{j\in\mathcal{K}\backslash \{k\}}{p_{j}}\left|\mathbf{u}^H\mathbf{G}_k\mathbf{h}\right|^2+\left\|\mathbf{z}^H\mathbf{G}_k\right\|^2\sigma_{r}^2+\sigma^2},\\
		&\qquad\qquad\; \textup{(42) -- (46), (48), (52), \eqref{P9_C1}, \eqref{P9_C2}},
	\end{align}
\end{subequations}
Problem \eqref{P10} is a convex optimization problem and can be efficiently solved by standard convex solvers, such as CVX.
\subsection{Proposed Algorithm and Computational Complexity}
\begin{algorithm}[!t]\label{method1}
	\caption{Proposed Low-Complexity Algorithm for Solving Problem \eqref{P2}}
	\begin{algorithmic}[1]
		\STATE {\bf Initialize} feasible points $\left\{\alpha_m^{(0)}, \mathbf{\Theta}^{(0)}, p_k^{(0)}\right\}$, energy efficiency $\xi^{(0)}=0$, and the maximum error tolerance $\epsilon_1$ and $\epsilon_2$.\\
		Set iteration index number $n_1=n_2=0$.
		\STATE {\bf repeat: outer loop}
		\STATE \quad {\bf repeat: inner loop}
		\STATE \quad\quad Obtain $\left\{\alpha_m^{(n_1+1)}, \mathbf{\Theta}^{(n_1+1)}\right\}$ by solving problem \eqref{P10} with given $\xi^{(n_2)}$ and $\{p_k^{(n_1)}\}$.\\
		\STATE \quad\quad Obtain $\{p_k^{(n_1+1)}\}$ by solving problem \eqref{P7} with given $\xi^{(n_2)}$ and $\left\{\alpha_m^{(n_1+1)}, \mathbf{\Theta}^{(n_1+1)}\right\}$.\\
		\STATE \quad\quad $n_1\leftarrow n_1+1$.
		\STATE \quad {\bf until} the fractional increase of the objective value is below $\epsilon_1$ or the maximum number of inner iterations is reached.
		\STATE \quad Update $\xi^{(n_2+1)}=\frac{\overline{R}{\left(\alpha_m^{(n_1)}, \mathbf{\Theta}^{(n_1)}, p_k^{(n_1)}\right)}}{P_{\rm tot}\left(\alpha_m^{(n_1)}, \mathbf{\Theta}^{(n_1)}, p_k^{(n_1)}\right)}$.	
		\STATE \quad $n_2\leftarrow n_2+1$, and $n_1\leftarrow 0$.
		\STATE {\bf until} $\left|\xi^{(n_2)}-\xi^{(n_2-1)}\right|\le\epsilon_2$ or the maximum number of outer iterations is reached.
		\STATE {\bf Output} the optimal solutions and maximum energy efficiency $\xi$.
	\end{algorithmic}
\end{algorithm}
Based on the two subproblems, i.e., problem \eqref{P10} and \eqref{P7}, the procedure for solving problem \eqref{P2} is summarized in \textbf{Algorithm 2}. The convergence analysis of \textbf{Algorithm 2} is similar to that of \textbf{Algorithm 1}. The details are omit here for brevity. Additionally, we claim that the obtained solution by the proposed algorithm may not necessarily be globally optimal. Then, the complexity of the proposed algorithm is analyzed as follows. Assuming the application of the interior-point method, the computational complexity for joint hybrid RIS optimization is $\mathcal{O}\left(2\left(KM^{3.5}\right)\right)$. Therefore, the overall computational complexity of \textbf{Algorithm 2} is ${{\mathcal{O}}}\left(I_{\rm out}I_{\rm inn}\left(2\left(KM^{3.5}\right)+K^{3.5}\right)\right)$, which is much lower than the proposed \textbf{Algorithm 1}. The effectiveness of the proposed low-complexity scheme will be further validated numerically in the next section.
\section{Numerical Results}
In this section, simulation results are provided to elaborate the effectiveness of the proposed hybrid RIS transmitter enabled energy-efficient multi-user communication system. 
\subsection{Simulation Setup}
\begin{figure}[t]
	\centering
	\setlength{\belowcaptionskip}{+0.1cm} 
	\includegraphics[width=	2.8in]{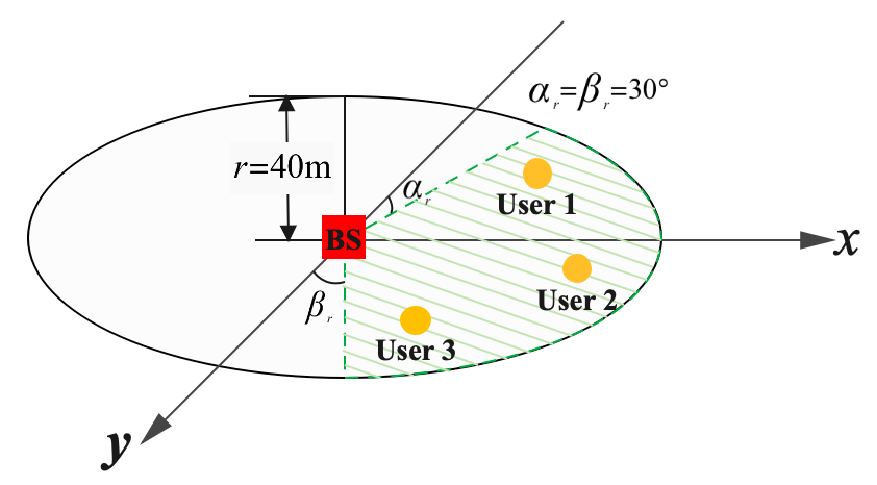}
	\caption{The simulated setup. }
	\label{Simulation setup}
\end{figure}
As Fig. 3 illustrated, the hybrid RIS is located at ($0$m, $0$m, $15$m). The users are randomly and uniformly distributed in a 120-degree sector region centered at ($0$m, $0$m, $0$m) with the radius of $r=40$m. For the number of RIS antennas, we set $M_x= 4$ and increase $M_y$ linearly with $M$. The antenna spacing is set to be half of the carrier wavelength, i.e., $\Delta_{t}=\frac{\lambda_c}{2}$. The carrier frequency is $f_c= 3$GHz.
Without loss of generality, we set $M = 20$, $P_{f}^{\rm max} = 20$dBm, $P_{r}^{\rm max} = 20$dBm, and $K=3$.  All users have identical minimum rate requirement, i.e., $R_k^{\rm min}=0.1$bit/s/Hz. The other simulation parameters are listed in Table I.
\begin{table}[t]
	\centering
	\caption{Simulation Parameters}\
	\begin{adjustbox}{width=3.5in}
		\begin{tabular}{|l|l|}
			\hline
			\multicolumn{1}{|l|}
			{\textbf{Parameter}} & \multicolumn{1}{l|}{\textbf{Value}}  \\ \hline
			Power amplifier efficiency & $0.8$~\cite{fang2022energy}   \\ 
			Power consumption of each RF chain & $48$mW~\cite{zi2016energy}  \\ 
			Power consumption of each RIS element & $6$dBm  \\ 
			Other circuit power consumption in the system & $30$dBm~\cite{liu2020energy}  \\ 
			Distance between the feed antenna and the hybrid RIS & $20$cm  \\ 
			Path-loss exponent of channels  & $2.2$  \\ 
			Path loss at the reference distance of 1 meter  & $-30$dB  \\ 
			Rician factor of the hybrid RIS-user channels  & $3$dB \ \\ 
			The noise power at the receivers and the hybrid RIS & $-60$dBm  \\ 
			Convergence accuracy   & ${10^{-3}}$   \\ \hline
		\end{tabular}
	\end{adjustbox}
	\centering
\end{table}

In the following, three baseline schemes are considered to compare with our proposed scheme:
\begin{itemize}
	\item 
	\textbf{Full-active RIS scheme}: In this case, the proposed algorithm is performed for energy-efficient design in a full-active RIS enabled multi-user network, in which all the transmitting elements operate in the active mode.
	The RIS mode scheduling matrix is set to be $\mathbf{A}\triangleq\textup{diag}({\mathbf{1}}_{1 \times M})\in \mathbb{B}^{M \times M}$.
	\item 
	\textbf{Full-passive RIS scheme}: In this case, the proposed algorithm is performed for energy-efficient design in a full-passive RIS enabled multi-user network, in which all the transmitting elements operate in the passive mode. The RIS mode scheduling matrix is set to be $\mathbf{A}\triangleq\textup{diag}({\mathbf{0}}_{1 \times M})\in \mathbb{B}^{M \times M}$.
	\item 
	\textbf{Coventional RF chain-based multi-antenna scheme}: 
	In this case, the conventional RF chain-based antenna design is adopted at the transmitter with a $M$-antenna ULA. We apply full-digital zero-forcing (ZF) precoding to suppress the interference caused by multi-user communication. The algorithms presented in this paper can be properly modified to achieve optimal power allocation design in the considered setup.	
\end{itemize}
\subsection{Convergence and CPU Running Time Comparison of the Proposed Algorithms}
\begin{figure}[t]
	\centering
	\setlength{\belowcaptionskip}{+0.1cm} 
	\includegraphics[width=	3.3in]{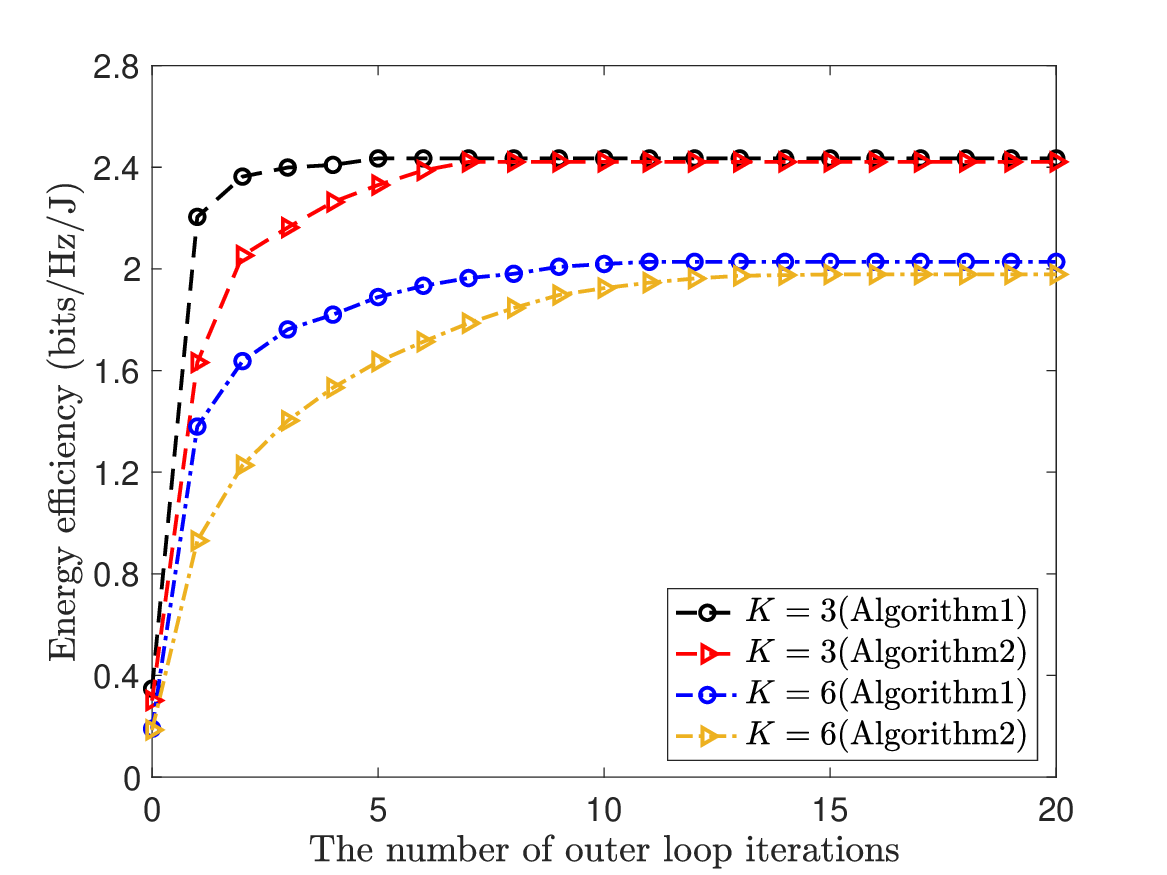}
	\caption{Convergence of the proposed two algorithms.}
	\label{Simulation results}
\end{figure}
\begin{table}[t]\small
	\centering
	\caption{CPU Running Time Required by the Proposed Algorithms}
	\begin{tabular}{c|c|c}
		\hline
		\textbf{Adopted Algorithm}   & \textbf{User Number} & \textbf{\begin{tabular}[c]{@{}c@{}}CPU Running Time \\ (Seconds)\end{tabular}} \\ \hline
		\multirow{2}{*}{Algorithm 1} & 3                    & 12657                                                                          \\ \cline{2-3} 
		& 6                    & 19824                                                                          \\ \hline
		\multirow{2}{*}{Algorithm 2} & 3                    & 34                                                                             \\ \cline{2-3} 
		& 6                    & 51                                                                            \\ \hline
	\end{tabular}
\end{table}
In Fig. 4, we first provide the convergence of both proposed algorithms over the same channel conditions with different numbers of users. It should be noted that \textbf{Algorithm 1} relies on an exhaustive search method.  A fairer comparison is to study the convergence for \textbf{Algorithm 1} with a given RIS mode scheduling scheme, in which case the system EE is maximized. As depicted in Fig. 4, the system EE of the proposed algorithms increases rapidly as the number of outer loop iterations progresses. Both algorithms converge within a few iterations$\footnote{Employing more advanced initialization schemes can substantially expedite convergence and improve the achievable performance of the developed algorithms.}$. The solution obtained by \textbf{Algorithm 1} serves as an upper bound performance for \textbf{Algorithm 2}. Additionally, the figure shows that an increasing number of users leads to lower system EE performance. With the growth in user count, the potential for interference among users rises, making interference management more complex.

Furthermore, Table II compares the average CPU running time used by each algorithm for execution. As shown in Table II, \textbf{Algorithm 1} requires much more average CPU running time compared to \textbf{Algorithm 2}. This is determined by the nature of exhaustive searching. Since an RIS usually consist of a large number of elements, there are numerous possibilities for index sets $\mathcal{A}$. In addition, the CPU running time of the algorithms is sensitive to the number of users. This is because the computational complexity of both algorithms is related to the value of $K$, which is consistent with our discussion in the previous sections.
\subsection{Impact of the Number of Transmitter Antennas}
\begin{figure}[t]
	\centering
	\setlength{\belowcaptionskip}{+0.1cm} 
	\includegraphics[width=	3.3in]{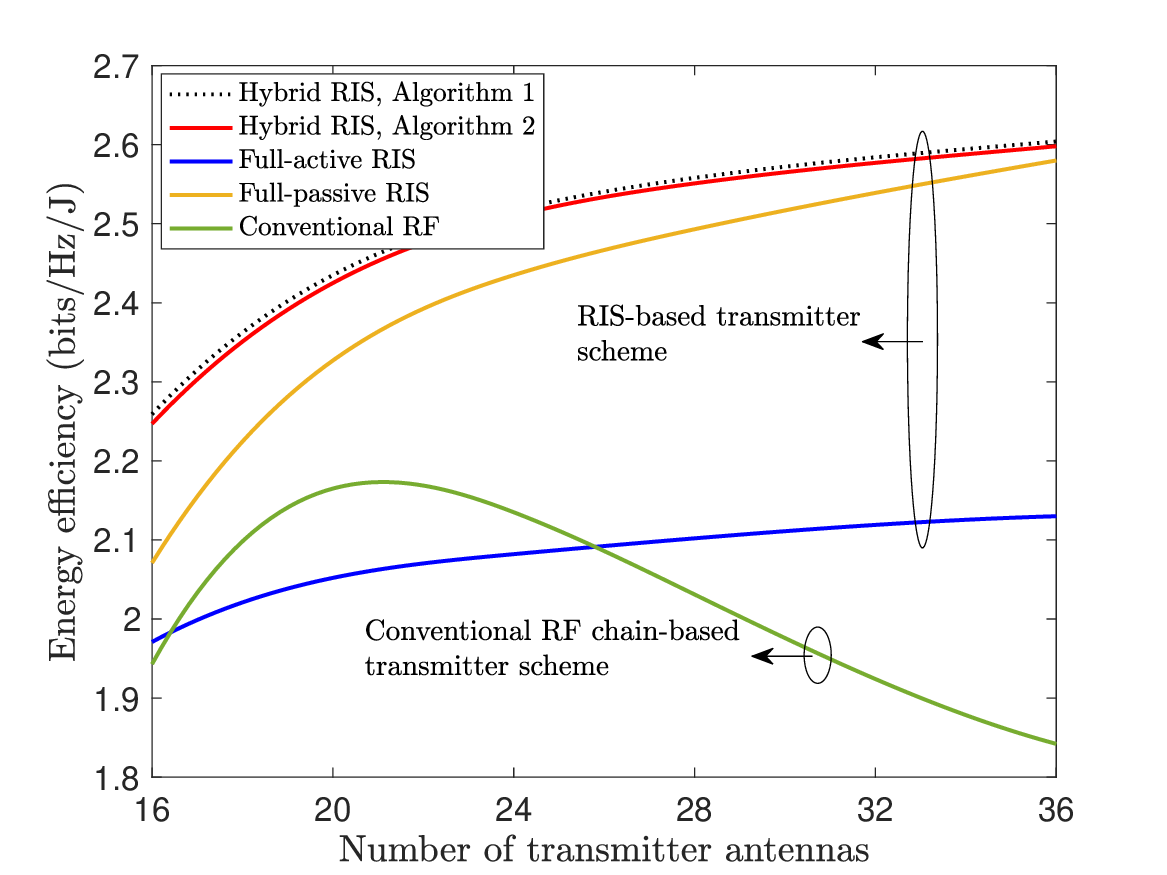}
	\caption{System EE versus the number of transmitter antennas $M$.}
	\label{Simulation results}
\end{figure}
Fig. 5 presents the system EE versus the number of RIS elements. As can be seen from the figure the system EE of all schemes increases with $M$. This is because a large number of RIS elements leads to a higher beamforming gain at a small power cost. Due to the flexibility in switching the element operating mode, i.e., passive or active, our proposed scheme is capable of achieving higher performance than other baseline schemes. We also observe that the performance gap between the solutions resulting from the two different algorithms is negligible, however, the computational complexity required by \textbf{Algorithm 1} is much higher than \textbf{Algorithm 2}. One more interesting finding is that as the antenna scale increases, the full-passive RIS scheme will gradually reap approximate performance to the proposed scheme. Benefiting from the high passive array gain in the large-size RIS case, it is possible to satisfy the minimum rate requirements of all users with fewer elements operating in an active mode. By activating only the necessary elements, the system can achieve better EE, which is essential for sustainable and cost-effective wireless communications. This figure demonstrates that our proposed scheme always outperforms the conventional RF chain-based transmitter scheme. RISs are free of energy-consuming RF chains and can significantly improve the system capacity with a reduced power budget. In addition, ZF precoding is effective in minimizing user interference, but at the expense of data rates, resulting in poorer performance gains. Furthermore, the figure confirms that there is an optimal number of transmitter antennas for maximizing EE. Due to the substantial power demand of each RF chain, there is a trade-off between the rate benefit of deploying larger array antennas and the associated energy consumption costs they incur.
\subsection{Impact of the RIS Mode Scheduling Scheme}
\begin{figure}[t]
	\centering
	\setlength{\belowcaptionskip}{+0.1cm} 
	\includegraphics[width=	3.3in]{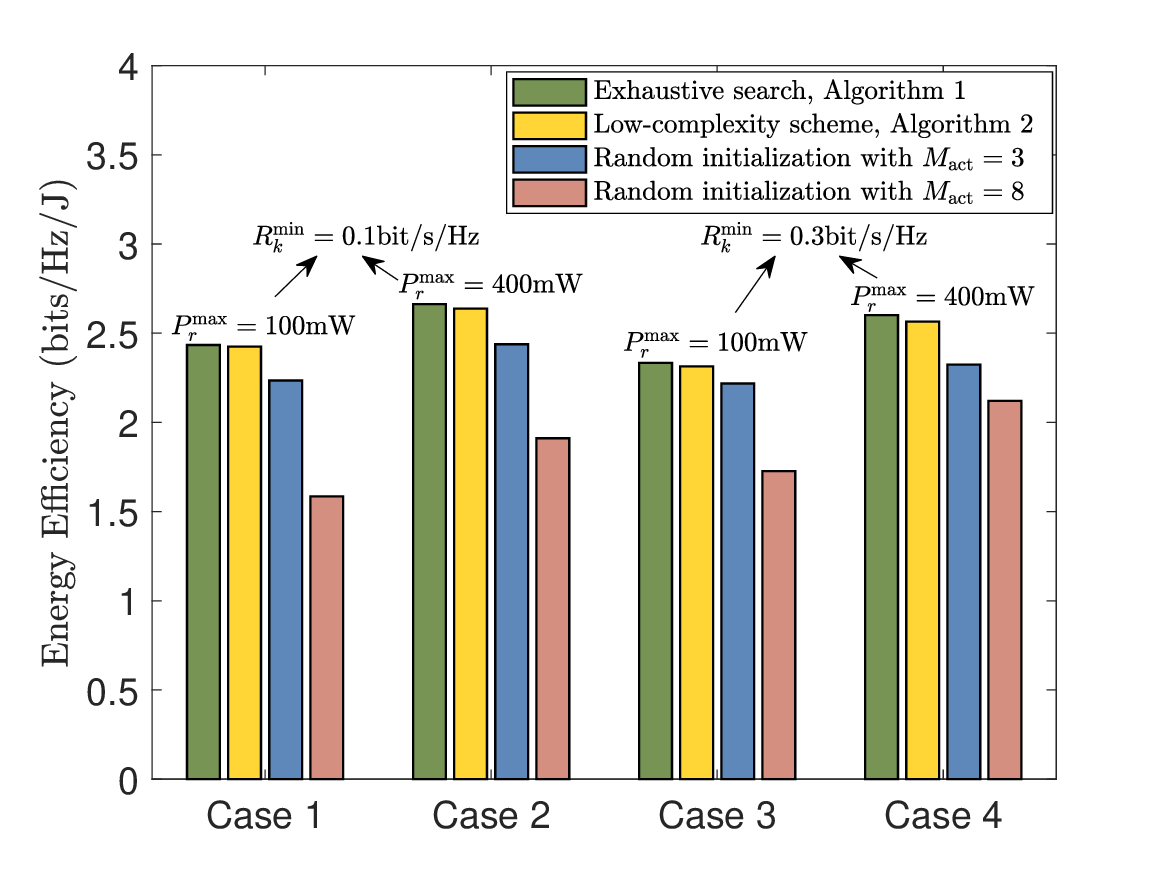}
	\caption{System EE versus different RIS mode scheduling schemes.}
	\label{Simulation results}
\end{figure}
Fig. 6 plots the system EE versus the different RIS mode scheduling schemes. The effectiveness of the two proposed algorithms for RIS element mode scheduling optimization is verified by comparing with the following baseline scheme:
\begin{itemize}
	\item 
	\textbf{Random initialization scheme}: In this case, the RIS element modes are randomly initialized with a defined number of active elements, i.e., the RIS mode scheduling matrix $\mathbf{A}$ is predefined without the need for optimization.
\end{itemize}
We consider four cases: (1) $P_r^{\max}=100$mW, $R_k^{\min}=0.1$bit/s/Hz; (2) $P_r^{\max}=400$mW, $R_k^{\min}=0.1$bit/s/Hz; (3) $P_r^{\max}=100$mW, $R_k^{\min}=0.3$bit/s/Hz; and (4) $P_r^{\max}=400$mW, $R_k^{\min}=0.3$bit/s/Hz. Additionally, in the baseline scheme, the hybrid RIS is assumed to comprise $3$ or $8$ active elements, respectively. The random initialization scheme results in a worse performance compared to the proposed schemes, which emphasizes the  importance of optimizing RIS element mode scheduling. We can see that by increasing $M_{\rm act}$, the EE performance of the system may be reduced instead. When more active RIS elements exist to share a limited power budget, the amplitude of each active element becomes smaller, causing signal attenuation and system performance degradation. In the considered setup, the proposed \textbf{Algorithm 2} is able to achieve a similar performance as \textbf{Algorithm 1} with relatively low-complexity, which is consistent with the conclusion obtained for Fig. 5. As a result, we can conclude that \textbf{Algorithm 2} is the preferred solution to optimize the RIS element mode scheduling in practice, and therefore we only evaluate the performance of this algorithm in the following subsections.
\subsection{Impact of the Minimum Rate Requirement of Users}
\begin{figure}[t]
	\centering
	\setlength{\belowcaptionskip}{+0.1cm} 
	\includegraphics[width=	3.3in]{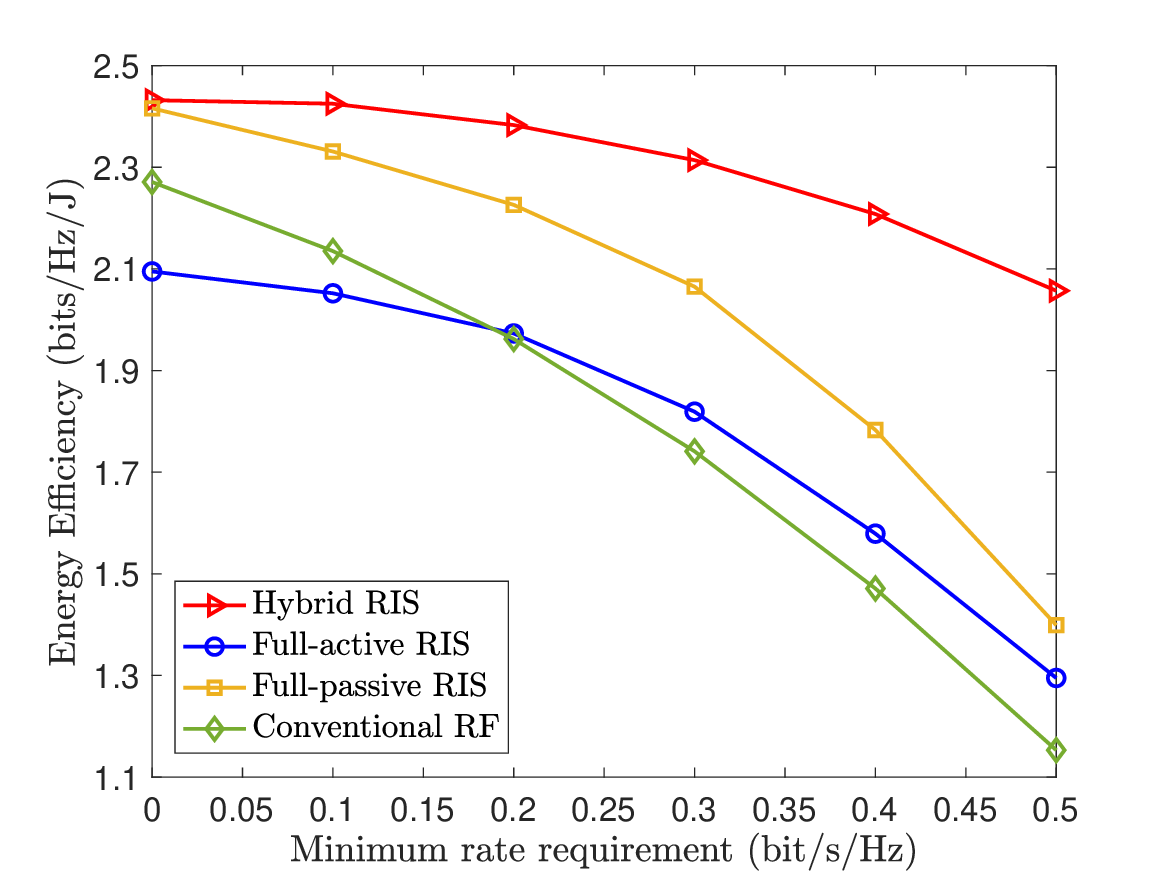}
	\caption{System EE versus the minimum rate requirement $R^{\min}_k$.}
	\label{Simulation results}
\end{figure}
Fig. 7 shows the system EE versus the minimum rate requirement of users. With a carefully designed operating mode of each element, there is always a noticeable performance gap between our proposed scheme and the baseline schemes using single-mode RISs. Additionally, leveraging the energy-saving feature of RISs, our proposed scheme notably surpasses conventional multi-antenna transmitter designs based on RF chains. As $R_k^{\min}$ increases, the performance gaps become more pronounced. This is due to the fact that our proposed scheme enables the system to well accommodate the high rate requirement, thereby ensuring only a marginal degradation in system performance. These findings demonstrate the potential of employing hybrid RISs for future transmitter antenna design, especially when there is an increasing demand for higher communication data rates.
\subsection{Impact of the Maximum RIS Amplification Power}
\begin{figure}[t]
	\centering
	\setlength{\belowcaptionskip}{+0.1cm} 
	\subfigure[System EE versus the maximum RIS amplification power.]{
	\includegraphics[width= 3.3in]{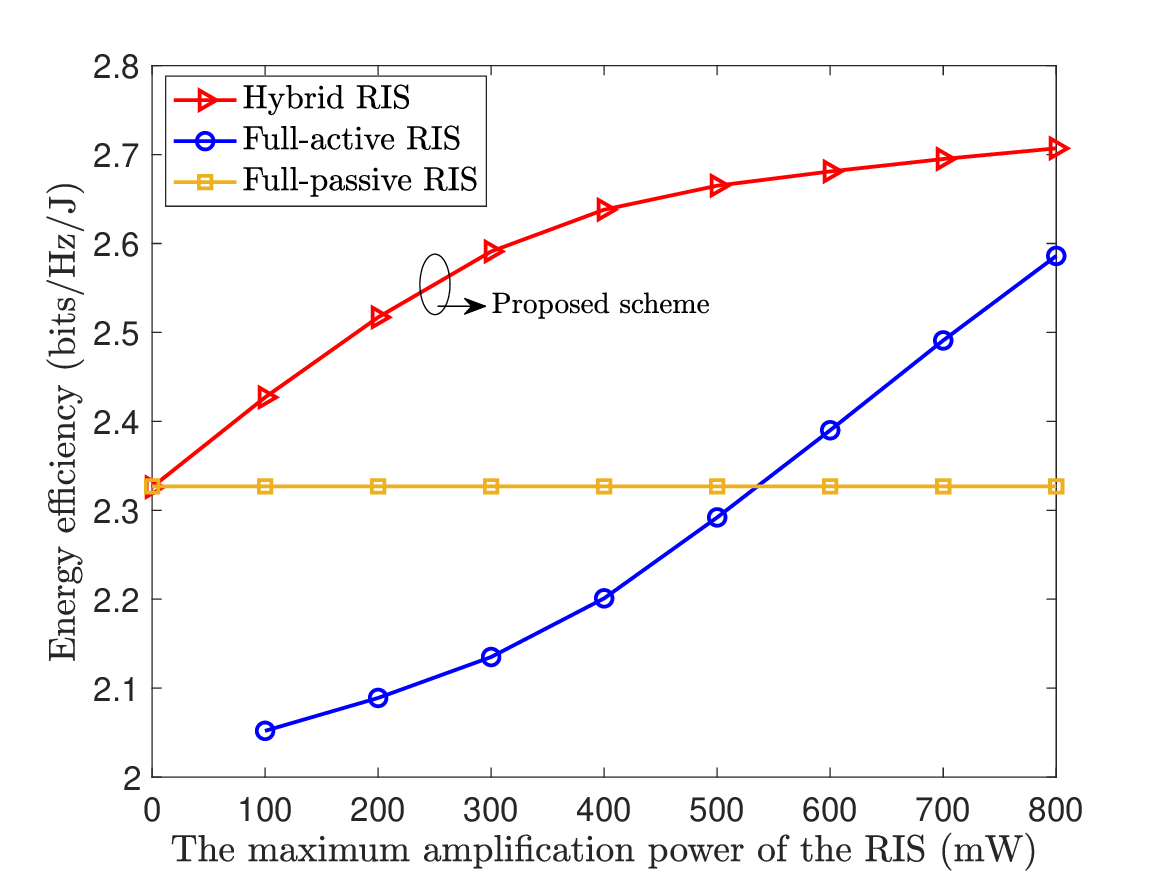}}
	\subfigure[The number of RIS passive/active elements versus the maximum RIS amplification power.]{
	\includegraphics[width=3.3in]{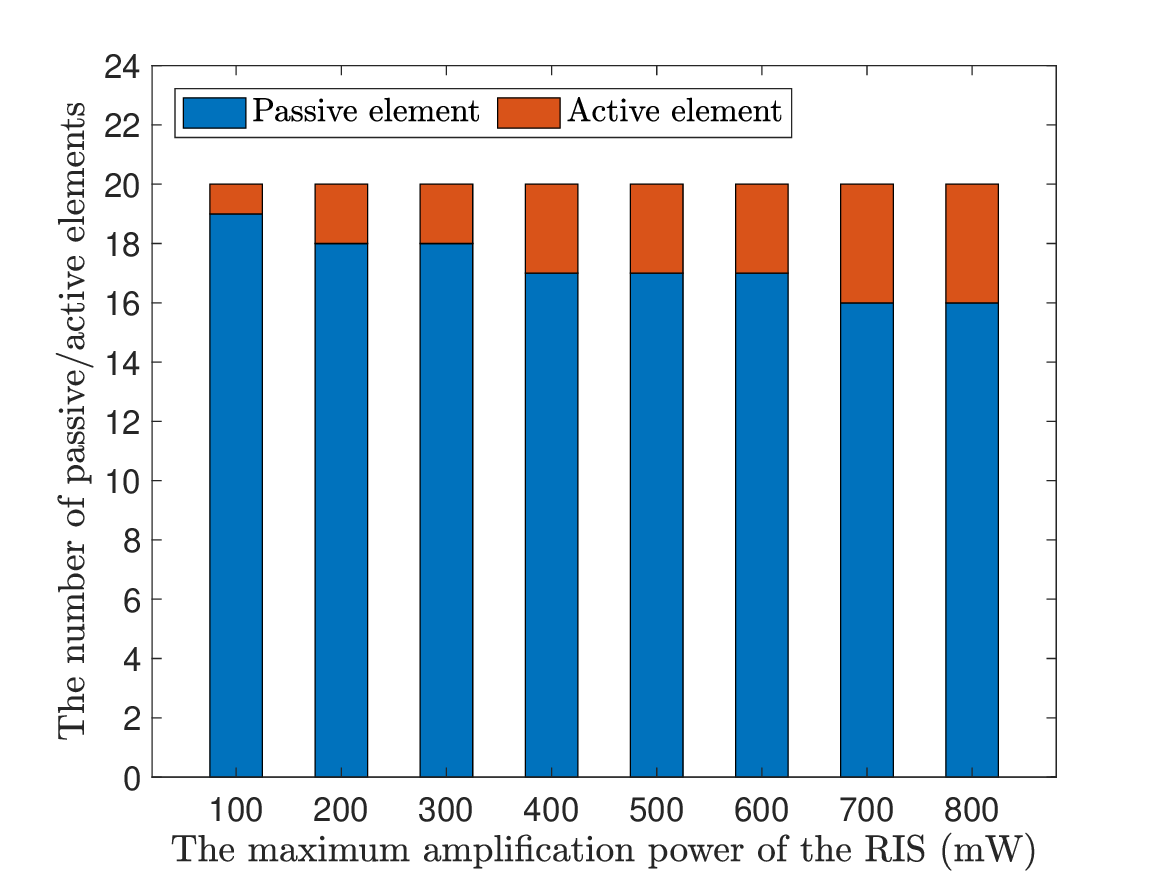}}
	\caption{Impact of the maximum RIS amplification power $P_{r}^{\rm max}$.}\label{Simulation results}
\end{figure}
In this subsection, we investigate the impact of the maximum RIS amplification power on system performance. Fig. 8(a) characterizes the system EE versus $P_{r}^{\rm max}$. The proposed scheme is significantly superior to the other two baseline schemes. Typically, the performance of schemes comprising active elements is greatly affected by the change of $P_{r}^{\rm max}$, whereas the full-passive scheme is not. Specifically, system EE of the proposed scheme first grow monotonically with the increase of $P_{r}^{\rm max}$ and then reach the maximum value. Due to the proper algorithm design, the hybrid RIS hardware design can effectively amplify the transmitted signal at a fraction of the power consumption. Nevertheless, there is an urgent need to pursue high amplification power to obtain maximum EE performance gain in full-active RIS enabled multi-user networks. Fig. 8(b) characterizes the number of passive/active elements versus $P_{r}^{\rm max}$. From this figure, we can observe that $M_{\rm act}$ is increased with the maximum power consumption allowed by the hybrid RIS, more elements are activated into the active mode to amplify the signal amplitude. This indicates that higher RIS amplification power makes an essential contribution to the enhancement of system performance. 
\section{Conclusions}
In this work, a novel downlink hybrid RIS transmitter enabled multi-user communication framework has been investigated, in which a hybrid RIS architecture is exploited to strike a trade-off between the performance benefits of signal amplitude modulation and power consumption. Based on the transmission design, a joint EE maximization problem was formulated. Efficient algorithms are developed to address the resulting MINLP problem by invoking Dinkelbach and AO techniques. Specifically, the optimal operating mode of each RIS element can be obtained by either the exhaustive search method or the proposed joint hybrid RIS optimization scheme. The latter permits the RIS element scheduling coefficients to be jointly optimized with different configurations of RIS amplitude and phase shifts. Our numerical results verified that the proposed scheme is superior to the baseline multi-antenna schemes employing fully active/passive RIS or conventional RF chains. The comparable performance can be achieved by the two proposed algorithms. Another interesting finding is that for a hybrid RIS, maximum EE can be achieved by setting only a few elements to operate in active mode, while increasing the number of active elements cannot further enhance communication performance. In the future, we will focus on investigating the fundamental trade-off between system SE and EE in hybrid RIS transmitter enabled networks, which is a promising direction that deserves in-depth research.
\balance
\bibliographystyle{IEEEtran}
\balance
\bibliography{mybib}

 \end{document}